\documentclass[aps,prl,reprint,longbibliography,superscriptaddress]{revtex4-1}

\usepackage{scrextend}

\usepackage{amsmath,bm}
\usepackage{mathrsfs}
\usepackage{amsfonts}
\usepackage{setspace} 
\usepackage{graphicx}
\usepackage{epstopdf}
\usepackage{dcolumn}
\usepackage{amsmath}
\usepackage{epsfig}
\usepackage{indentfirst}
\usepackage{psfrag}
\usepackage{subfigure}
\usepackage{amssymb}
\usepackage{color}
\usepackage{xcolor}
\usepackage{siunitx}
\usepackage{lipsum}
\usepackage{graphicx}
\usepackage{dcolumn}
\usepackage{bm}
\usepackage{physics}

\usepackage[backref=none,bookmarksnumbered=true,bookmarks=true,bookmarksopen=true,colorlinks=true,
citecolor=blue,linkcolor=blue,anchorcolor=green,urlcolor=blue,unicode=false]{hyperref}

\usepackage{ulem}[normalem] 

\usepackage{titlesec}
\titleformat{\section}[block]{\normalfont\bfseries\centering}{\thesection}{1pt}{}

\makeatletter
\newcommand\colorsout[1]{\bgroup \markoverwith{\textcolor{#1}{\rule[0.5ex]{2pt}{0.4pt}}}\ULon}

\makeatother

\AtBeginDocument{%
    \newwrite\bibnotes
    \def\bibnotesext{Notes.bib}
    \immediate\openout\bibnotes=\jobname\bibnotesext
    \immediate\write\bibnotes{@CONTROL{REVTEX41Control}}
    \immediate\write\bibnotes{@CONTROL{%
    apsrev41Control,author="08",editor="1",pages="1",title="0",year="1"}}
     \if@filesw
     \immediate\write\@auxout{\string\citation{apsrev41Control}}%
    \fi
}%

\definecolor{orange}{rgb}{1,0.5,0}
\definecolor{goodgreen}{rgb}{0.1,0.5,0}
\definecolor{goodred}{rgb}{0.7,0,0}
\let\oldepsilon\epsilon \let\epsilon\varepsilon \let\varepsilon\oldepsilon

\makeatother

\usepackage{babel}

\begin{document}

\title{From local to collective superconductivity in proximitized graphene}

\author{Stefano Trivini}

\affiliation{CIC nanoGUNE-BRTA, Donostia-San Sebasti\'an, Spain}
\affiliation{Centro de Física de Materiales (CFM-MPC), Donostia-San Sebasti\'an, Spain}

\author{Tim Kokkeler}
  \affiliation{Donostia International Physics Center (DIPC), Donostia-San Sebastian, Spain}
\affiliation{University of Twente, Enschede, The Netherlands}

\author{Jon Ortuzar}
\affiliation{CIC nanoGUNE-BRTA, Donostia-San Sebasti\'an, Spain}

\author{Eva Cortés-del Río}
\affiliation{Departamento de Física de la Materia Condensada, UAM, Madrid, Spain}
\affiliation{Department of Physics, University of Hamburg, Germany}

\author{Beatriz Viña Baus\'a}
\affiliation{CIC nanoGUNE-BRTA, Donostia-San Sebasti\'an, Spain}
\affiliation{Departamento de Física de la Materia Condensada, UAM, Madrid, Spain}

\author{Pierre Mallet}
\affiliation{Université Grenoble Alpes, CNRS, Institut Néel Grenoble, France}

\author{Jean-Yves Veuillen}
\affiliation{Université Grenoble Alpes, CNRS, Institut Néel Grenoble, France}

\author{Juan Carlos Cuevas}
\affiliation{Dpto. de F\'{\i}sica Te\'orica de la Materia Condensada, UAM, Madrid, Spain}

\affiliation{Condensed Matter Physics Center (IFIMAC), UAM, Madrid, Spain}

\author{Ivan Brihuega}
\affiliation{Departamento de Física de la Materia Condensada, UAM, Madrid, Spain}

\author{F. Sebastian Bergeret}
 \affiliation{Centro de Física de Materiales (CFM-MPC), Donostia-San Sebasti\'an, Spain}
\affiliation{Donostia International Physics Center (DIPC), Donostia-San Sebastian, Spain}

\author{Jose Ignacio Pascual}
  \affiliation{CIC nanoGUNE-BRTA, Donostia-San Sebasti\'an, Spain}
\affiliation{Ikerbasque, Basque Foundation for Science, Bilbao, Spain}

\begin{abstract}  
The superconducting proximity effect induces pairing correlations in metallic systems via Andreev scattering. This effect is particularly intriguing in graphene, as it enables two-dimensional superconductivity that is tunable through doping. Understanding how superconducting correlations propagate within the metal is crucial to unveiling the key factors behind this tunability. Here, we employ scanning tunneling microscopy to investigate the energy and length scales of the proximity effect induced by Pb islands on graphene. Using tip-induced manipulation, we assemble S/N/S junctions with tunable N-region spacing and explore the evolution of the proximitized state in the confined normal region. We find that different doping levels can lead to either localized or collective superconducting states. By combining our experimental results with quasiclassical theory, we demonstrate that interface conductance plays a key role in determining the strength and coherence length of pairing correlations and inter-island coupling. Our findings provide new insights into the design of novel superconducting states and the control of their properties.
\end{abstract}

\maketitle


Superconducting pairing can be induced in normal metals (N) that are in good contact with a superconductor (S), where Andreev reflection governs electronic transfer across the interface \cite{andreev1964,klapwijk2004}. This proximity effect is essential for inducing superconductivity in quantum materials, enabling a wide range of novel phases of matter and exotic effects, including unconventional pairing \cite{bergeret2005,black-schaffer2013,cao2018,balents2020}, superconducting topological states \cite{leijnse2012,san-jose2015,frolov2020}, non-reciprocal supercurrents \cite{ando2020,ilicc2024}, and many-body correlations in magnetic impurities \cite{kezilebieke2020b,manna2020a,stolyarov2021,vaxevani2022,trivini2023a}.

The length scales of superconducting pairing propagation into a normal (N) metal are crucial for transforming proximitized regions into a macroscopic superconducting state.  In a normal region, the coherence length  $\xi(E)$ can be much larger than the electron mean free path, justifying a diffusive description of the proximity effect \cite{usadel1970}. It also depends inversely on the correlated electron/hole energy $E$,  following the relation $\xi(E) = \sqrt{\hbar D/E}$, where $D$ is the diffusion coefficient.  Consequently, the  normal local density of states (LDoS) in N is depleted over length scales ranging from  $\xi(\Delta_S) = \sqrt{\hbar D/\Delta_S}$ at $E=\Delta_S$, the superconductor order parameter, to longer distances for $E \sim $ 0.  If N has a finite length $d$, the proximity effect induces a hard (mini)gap $\Delta_{\text{p}} \leq \Delta_S$, which is primarily determined by the Thouless energy  $ E_{\text{Th}} = \hbar D / d^2\ \label{thouless}$ \cite{fominov2001superconductive,golubov2004the,giazotto2010superconducting,cherkez2014proximity,hijano2021coexistence,strambini2017revealing,kashiwaya1998tunneling,kashiwaya1995origin}. 

\begin{figure}[b]
\begin{center}
    \includegraphics[width=1\columnwidth]{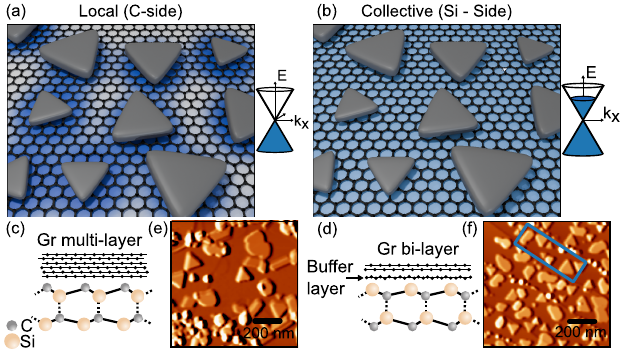}    
	\caption{\textbf{Local and collective proximity in SiC graphene.}
(a,b) Strength and distribution of superconductivity (in blue) induced by lead islands on graphene over the two sides of a SiC crystal with different native doping.  Crystal structures of the (c) charge neutral graphene multilayer on the C-side, and (d) the n-doped graphene bi-layer on the Si side of a SiC crystal. 
(e,f) STM images of nanoscale Pb islands grown on graphene over the C-side and Si-side of a SiC  crystal \cite{horcas2007}. Tunnel parameters: V = \SI{100}{mV}, I = \SI{100}{pA}. } \label{F1}
  \end{center}
\end{figure}

\begin{figure*}[t]
    			\includegraphics[width=0.99\textwidth]{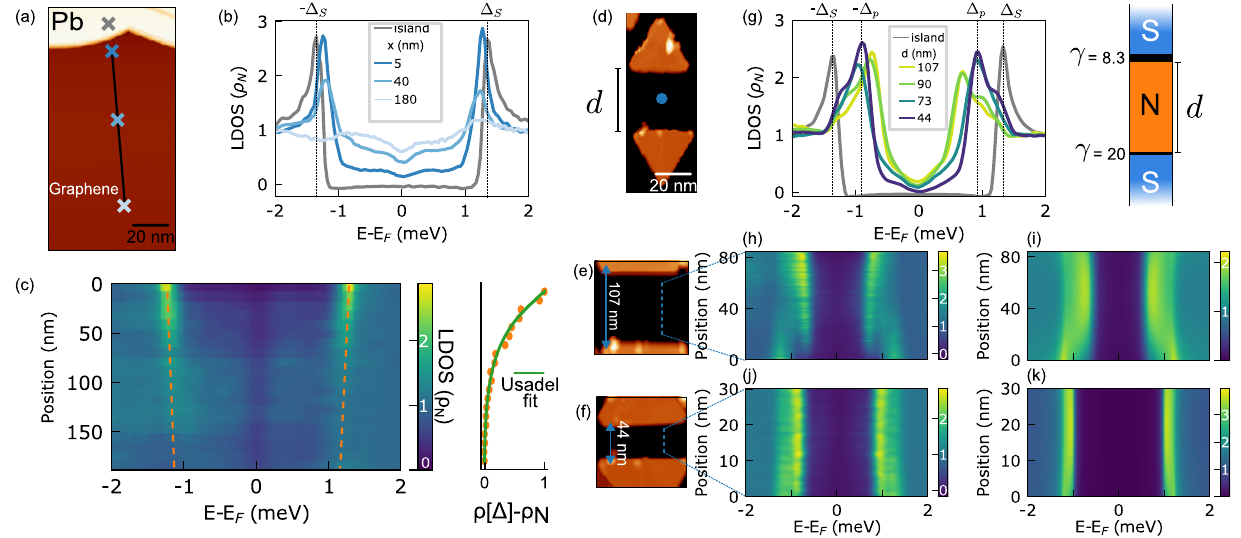}    
			\caption{\textbf{Proximitized superconductivity on C side graphene.} 
    (a) STM image of a Pb island on top of C-side graphene. Line and crosses refer to the points at which the \textit{dI/dV} spectroscopy in (b) and (c) are taken. 
    (b) Deconvoluted \textit{dI/dV} spectra (LDoS) measured on the crosses in (a).  
   (c) Deconvoluted spatial \textit{dI/dV} along the line in (a). Right, decay of the coherent peaks' amplitude with distance, fitted using Usadel equations for a N/S junction using  $\xi_S$ = \SI{200}{nm}, and  $\Gamma = 0.04\,\Delta$.
   (d) STM image of an S/N/S junction built by lateral manipulation of two islands.
   (e) Zoom in of the same junction at 107 nm N-region.
   (f) Same junction after further STM-tip manipulation to 44 nm N-region.
   (g) Deconvoluted \textit{dI/dV} spectra at the center of the S/N/S junction for different island separations.
    (h) Deconvoluted spatial \textit{dI/dV} in the \SI{107}{nm} N-region.
    (i) Simulated LDoS of \SI{107}{nm} S/N/S, with    $\Delta =$ \SI{1.35}{mV} $\Gamma = 0.04\,\Delta$, $\xi_S =$ \SI{140}{nm}, $d=$ \SI{83}{nm}, top electrode $\gamma = 8.3$, bottom electrode $\gamma=20$. 
    (j) Deconvoluted spatial \textit{dI/dV} in  \SI{44}{nm} long N-region.  
    (k) Simulated LDoS of the \SI{44}{nm} S/N/S. Parameters: $\Delta =$ \SI{1.35}{mV}, $\Gamma = 0.04\,\Delta$, $\xi_S =$ \SI{140}{nm}, $d=$ \SI{30}{nm}, top electrode $\gamma = 20$, bottom electrode $\gamma=8.3$. STM topographies at V = \SI{100}{mV}, I = \SI{100}{pA}, \textit{dI/dV} at V = \SI{5}{mV}, I = \SI{500}{pA} with $\delta V=$~100~uV lock-in modulation.
    Deconvolution with $\Delta_t =$ \SI{1.35}{mV}.}
		\label{F2}
\end{figure*}

However, the efficiency of superconducting proximity is often limited by interface properties, frequently modeled by a transparency parameter that describes the balance between conventional electron scattering and Andreev reflection  \cite{blonder1982a}. The interface transparency, usually parametrized by a dimensionless interface conductance ratio $\gamma = G_I / G_N$  of interface (I) and N \cite{kuprianov1988,belzig1996,hammer2007},  plays a key role in determining the strength of pairing correlations in N and their characteristic length scales. Reducing the values of $\gamma$ causes the narrowing of $\Delta_{\text{p}}$ and larger $\xi(\Delta_{\text{p}})$ lengths \cite{aminov1996a}. 
Thus, understanding the role of the N/S interface in the proximity effect requires determining the size and shape of the proximity gap, as well as its dependence on length \cite{gueron1996}.   While scanning tunneling microscopy (STM) techniques have improved our ability to map the spatial dependence of the proximity-induced gap \cite{lesueur2008,natterer2016,stolyarov2021,cortes-delrio2021,cortes-delrio2023}, the experimental determination of $\gamma$ remains elusive.

Here, we demonstrate that the strength and extension of the superconducting proximity effect in graphene can be controlled by the transparency of the superconductor-normal (S/N) interface. We compare STM measurements around Pb islands (S) grown on graphene (N) on the C- vs.\ Si-terminated sides of a SiC crystal and observe striking differences in the induced superconducting state, as summarized in Fig.~\ref{F1}a and \ref{F1}b. On the C-side, the proximity effect is locally strong but spatially inhomogeneous, fading with distance from the islands; on the Si-side, the induced superconducting gap $\Delta_{\text{p}}$ is smaller yet extends homogeneously throughout the graphene layer, revealing a collective superconducting state. Through quasi-classical modeling of the proximity effect \cite{larkin1969quasiclassical,eilenberger1968transformation,usadel1970,golubov1988,virtanen2007}, we attribute these different behaviors to variations in the S/N interface conductance $\gamma$. In particular, the weaker superconducting state induced on the Si-side shows longer coherence lengths, favoring a uniform proximity gap collectively determined by many islands. These findings establish the S/N interface conductance as a key parameter governing superconductivity in graphene-based hybrid systems.

\begin{figure*}[t]
\begin{center}
    			\includegraphics[width=0.99\textwidth]{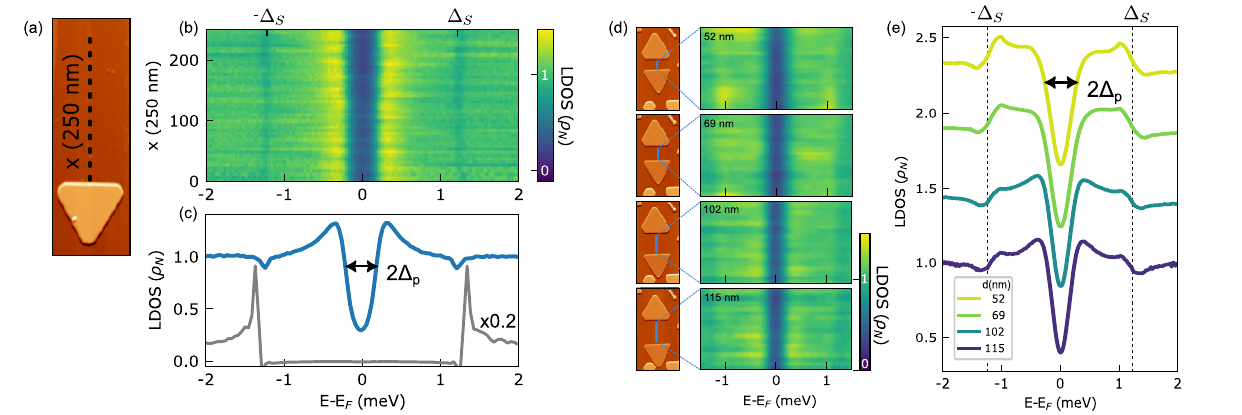}    
			\caption{\textbf{Proximity induced superconductivity on Si side graphene.}
    (a) STM image of the region in Fig.~\ref{F1}h with all islands removed but the shown one. (b) Spatial dependence of the LDoS in Si-side graphene along the dashed line in (a). (c) Deconvoluted \textit{dI/dV} spectra measured on top of (gray, normalized for clarity) and near to (blue) the island shown in panel (a).  The minigap amounts to $\Delta_{\text{p}} =$ \SI{0.2}{mV}. (d) Four snapshots of the closing of an S/N/S junction and deconvoluted \textit{dI/dV} spectral profiles along the N region (blue lines in the images). (e) Deconvoluted dI/dV curves in the center or the S/N/S junction in (d).  STM with V = 100mV and I = 20pA; \textit{dI/dV} setpoint: V = 6~mV, I = 600 pA. Deconvolution with $\Delta_t =$ \SI{1.35}{mV}.}
		\label{F3}
  \end{center}
		\end{figure*}
    
\textit{Preparation of the lead/graphene system.---} 
Graphene layers can grow on the two surfaces of hexagonal silicon carbide crystals by annealing it to high temperatures \cite{norimatsu2014}. 
As we show in the Supplemental Material (SM \cite{SI}), graphene grows as as charge-neutral multilayer on the C-terminated SiC(000$\overline{1}$) side (\textit{C-side}) and as an electron-doped bilayer on the Si-terminated SiC(0001), the \textit{Si-side} [Figs.~\ref{F1}c and \ref{F1}d] \cite{luxmi2010}. 
In two different experiments, we deposited lead ($\text{T}_c =$ \SI{7.2}{K}) by thermal sublimation on graphene on different sides of SiC crystals, at room temperature and under ultra-high vacuum conditions. On both substrates, lead forms polygonal islands with lateral sizes of $\sim$ 50-200~nm and height of  $\sim$ 3-10~nm (Fig.~\ref{F1}e-\ref{F1}f). 
To study the effect of Pb islands on the different graphene surfaces, we used a low-temperature scanning tunneling microscope at 1.2 K. We acquired differential conductance dI/dV spectra using Pb superconducting tips \cite{rodrigo2004,franke2011a}. All spectra reported are deconvoluted to obtain the sample LDoS \cite{pillet2010a}. The Pb islands display a bulk-like BCS gap with two coherence peaks at $\Delta_S=$ \SI{1.35}{mV}, characteristic of Pb(111) \cite{ruby2015}, broadened by an inelastic term $\Gamma=0.04~\Delta_S$.


\textit{Proximity effect of graphene on C-side SiC.---}
We first investigate the proximity effect induced by Pb islands on the C-side graphene [Fig.~\ref{F2}a], focusing on moiré-free domains that behave like free-standing graphene \cite{hass2008,sprinkle2009}.   Spectra acquired near Pb islands displays pronounced coherence peaks near $\pm \Delta_S$  [Fig.~\ref{F2}b],  which gradually diminish with distance from the islands [Fig.~\ref{F2}c]. Such evolution of the LDoS in the N-region is well-described by a one-dimensional quasiclassical model of an S/N system with an infinite N region, as detailed in \cite{virtanen2007}. Applying the model to reproduce the spectral decay in Fig.~\ref{F2}c, we obtain a coherence length $\xi_S = \xi(\Delta_S) \approx 200~\pm $  \SI{10}{nm}, consistent with previous reports \cite{natterer2016}, and a diffusion coefficient $D\sim$ \SI{820}{cm^2/s}.

Graphene regions like the one in Fig.~\ref{F2}a exhibit a net conductance for all sub-gap energies [Fig.~\ref{F2}b]  indicating incomplete proximitization, as expected for an infinite N-region \cite{cherkez2014proximity}. Surprisingly, graphene domain boundaries on the C-side, typically spaced with distances of a few times $\xi_S$ \cite{varchon2008}, do not induce mini-gaps inside the coherence peaks,    
but at most to faint zero-energy LDoS dips [Fig.~\ref{F2}b], probably remnant of a narrow gap smeared by finite temperature (\SI{1.2}{K}), or additional pair-breaking mechanisms.

To induce a mini-gap in graphene, we confined N-regions between two Pb islands to lengths comparable to $\xi_S$. The S/N/S junction [Fig.~\ref{F2}d] was created through STM tip manipulation \cite{eigler1990,schneider2023,rutten2024a,trivini2024b} by laterally dragging triangular Pb islands to the desired positions \cite{cortes-delrio2023}. Two Pb islands were initially positioned face to face on a moiré-free graphene region, separated by 107 nm [Fig.~\ref{F2}e]. The spacing was gradually reduced to 44 nm by moving one island closer to the other using an STM tip [Fig.~\ref{F2}f]. The spectral LDoS taken in the middle between the islands at each manipulation stage of 107, 90 , 73 and 44 nm [Fig.~\ref{F2}g] reveals the emergence of a proximity gap varying from $\sim$ \SI{0.6}{meV} to \SI{0.9}{meV} and the in-gap conductance gradually decreasing.

The wider S/N/S junction exhibits a spatially asymmetric LDoS [Fig.\ref{F2}h] with distinct proximity-induced gap values near the upper and lower islands. Since both islands have the same superconducting gap $\Delta_S$, and the Thouless energy $E_{\text{Th}}$ is defined by the N region size, the asymmetry likely arise from variations in the transparency of the island-graphene interfaces. As shown in Fig.~\ref{F2}i, this behavior is well captured by the 1D quasiclassical model, assuming a lower interface conductance for the top island ($\gamma=8.3$) compared to the bottom one ($\gamma=20$).

In contrast, the shorter S/N/S junction reveals a wider and more uniform proximity-induced gap along the N region [Fig.~\ref{F2}j], with $\Delta_{\text{p}} \sim$ \SI{0.9}{meV}. The model in Fig.~\ref{F2}k captures both the wider gap, consistent with a higher $E_{\text{Th}}$, and its homogeneity, even when using the same $\gamma$ values as for the more extended junction. This uniformity arises because the N region is much shorter than the superconducting coherence length. These results highlight the strong distance dependence of the proximity effect in graphene and the critical role of interface conductance in shaping proximity-induced mini-gaps.


\textit{Proximity effect of graphene on Si-side SiC.---} 
On the Si-side of a SiC crystal, graphene proximitized by Pb islands — of similar size and number as those on the C-side — exhibits notable differences. The most striking is the presence of a pronounced mini-gap, homogeneously distributed all over the graphene sample.  To illustrate this, we removed the surrounding islands around the selected region highlighted in Fig.~\ref{F1}f. Probing the spectral evolution with distance from the selected island [Fig.~\ref{F3}a], we remarkably find that the proximity gap remains spatially uniform over 250 nm [Fig.~\ref{F3}b], in contrast to the local variations observed on the C-side in Fig.~\ref{F2}c. The mini-gap has a width $\Delta_{\text{p}} \sim$ \SI{0.2}{meV} [Fig.~\ref{F3}c], with faint dips at $\Delta_S$ = \SI{1.3}{meV}, reminiscent of Pb island coherence peaks. 

In contrast to the C-side, the LDoS in regions confined between two islands is relatively unperturbed. In the \textit{tunable} S/N/S junction shown in Fig. \ref{F3}d, we compare the LDoS in the N region for island separations $d$ $=$ 115,  102, 69, \SI{52}{nm}, and find that it stays nearly constant despite varying $d$. 
Even for separations below the reported $\xi_S\approx$ \SI{150}{nm}, rescaled to $\Delta_S$ from Ref. \cite{natterer2016}, the proximity gap retains the same width $\Delta_{\text{p}}$ as in other regions. The confinement only induces a slight increase in the LDoS near $\Delta_S$, as visible in the average spectra in Fig.~\ref{F3}e. 

The robustness of $\Delta_p$ upon island removal and its homogeneity around the sample suggests that, on the Si-side, the LDoS is not determined by individual island proximitization, but rather by a collective effect involving all Pb islands over extended graphene regions, which stabilizes a global proximity gap \cite{feigelman2008}. Unraveling the origin of such differences requires extending the model to this configuration.


\begin{figure}[t]
\begin{center}
    			\includegraphics[width=1\columnwidth]{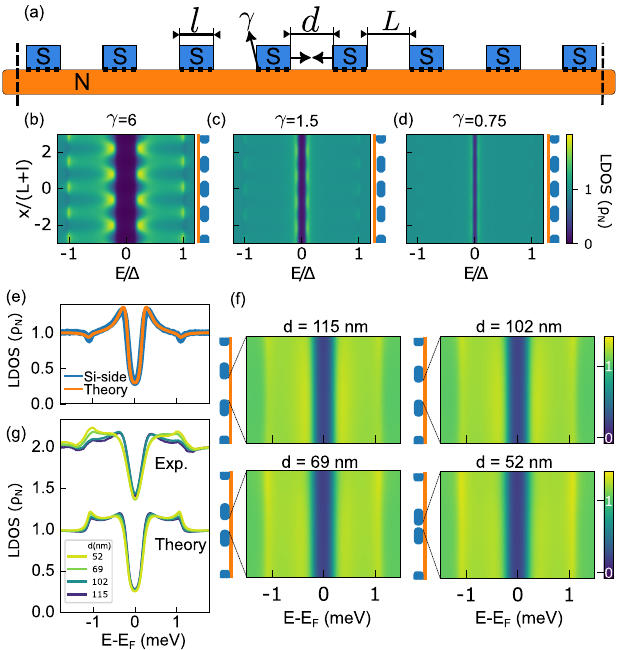}    
			\caption{ \textbf{Simulation of a periodic array of islands.}
    (a) Sketch of model used to simulate the proximity effect on the Si side, composed of a slab of normal metal with an array of islands in contact with it, extended with periodic boundary conditions.
    (b,c,d)  LDoS along N slab for different interface conductance values  $\gamma=6,~1.5,~0.75$, using $\xi_S=$ \SI{150}{nm}, $L=d=$ \SI{300}{nm}, $l=$ \SI{150}{nm}, and $\Gamma=0.04~\Delta_S$.
    (e) The experimental LDoS on the Si-side from Fig.~\ref{F3}b can be reproduced with the array model using parameters $\xi_{S}=$ \SI{150}{nm}, $L=$ \SI{375}{nm}, $l =$ \SI{187.5}{nm}, $\gamma = 4.4$.
    (f) Calculated  LDoS between two islands with different spacings $d = $115, 102, 69, \SI{52}{nm}, keeping the remaining array in the unit cell with periodicity $L=$ \SI{106}{nm} and $l=$ \SI{53}{nm}, and using $\xi_S=$ \SI{150}{nm},   $\gamma=1$ and $\Gamma=0.04\Delta_S$.
    (g) Calculated LDoS at the mid-point between the two central leads spaced by  $d$ in (f), qualitatively reproducing the experimental spectra presented in Fig.~\ref{F3}d.}
		\label{F4}
  \end{center}
		\end{figure}

\textit{Modeling the collective proximity effect.---} 
The one-dimensional model used above cannot capture the collective proximity effect induced by multiple islands. Therefore, we adopt a two-dimensional configuration, shown in Fig.~\ref{F4}a (see \cite{SI} for more details on the model), to solve the quasiclassical equations. 
The system consists of an infinite normal slab (N) proximitized by an array of superconducting islands (S) placed on top, with interface conductance $\gamma$, and dimensions extracted from the experiments: island size $l = $ \SI{53}{nm}, spacing $L = $ \SI{106}{nm}, and superconducting coherence length $\xi_S = $ \SI{150}{nm} \cite{natterer2016}.

For interface conductance $\gamma = 6$, the calculated LDoS along the N-slab shows a proximity-induced gap with strong spatial variations between the islands [Fig.~\ref{F4}b], resembling the behavior on the C-side. Reducing $\gamma$ yields smaller, more uniform gaps [Fig.~\ref{F4}c], ultimately forming a nearly homogeneous mini-gap at $\gamma = 0.75$, with LDoS features at $E\sim\Delta_S$ washed out. 
This scenario seemingly reproduces the collective proximitization observed on the Si-side. As shown in Fig.~\ref{F4}e, the model captures the LDoS features of the measured spectra in this multi-island regime.
  
The most striking feature of the Si-side proximity effect is the stability of the gap against confinement in an S/N/S junction [Fig.~\ref{F3}d-e]. This behavior is not captured by the 1D model, which predicts gap opening for any $\gamma$, but is well explained by the array model. Simulating a tunable S/N/S junction with N-region length $d$ embedded in a periodic array of Pb islands with spacing $L$, we reproduce the stability of the LDoS on $d$ [Fig.~\ref{F4}f]. As in the experiment, the proximity gap remains constant despite varying $d$. 

The array model also captures enhanced spectral weight at $\Delta_S$, as seen in the simulated LDoS spectra in Fig.~\ref{F4}g. Furthermore, the collective nature of the superconducting state on the Si-side is confirmed by selectively removing Pb islands from the array, as demonstrated in the SFig.~5 in SM \cite{SI}. This perturbation produces only slight changes in the proximity gap $\Delta_{\text{p}}$, an effect accurately reproduced by the model.

\textit{Discussion.---} 
On the C-side, our results point to highly transparent interfaces, leading to intense coherence peaks at $\Delta_S$ and a large proximity-induced gap, which strongly depends on the size of the normal region. This aligns with the inverse dependence of the Thouless energy on the size of N, giving $\xi_S\sim$ \SI{200}{nm} at  $E=\Delta_S=$ \SI{1.35}{mV}.
In constrast, the Si-side shows a small, uniform proximity gap $\Delta_p\approx$\SI{0.2}{meV}, about one-sixth of $\Delta_S$, stable across the sample and in confined regions. The difference is attributed to significantly less transparent Pb-graphene interfaces. 

The homogeneous LDoS on the Si-side reveals a collective superconducting state proximitized in the graphene layer. From the equation $\xi(E) = \sqrt{\hbar D/E}$, a smaller gap due to a more opaque interface results in a longer coherence length  $\xi_{\Delta_{\text{p}}}$. In fact, from the value $\Delta_p $  $ \sim $ \SI{600}{nm}, rescaled to $\Delta_p$ from \cite{natterer2016}—about six times larger than the average inter-island distance of \SI{100}{nm}. Consequently, several tens of  Pb islands can  be proximity-coupled for the formation of a collective superconducting state, even without a periodic arrangement \cite{kessler2010a,feigelman2008}. This results in an homogeneous spectral mini-gap across the sample.
Modeling graphene as a single object proximitized by multiple islands captures the observed stability and uniformity of $\Delta_p$ — a hallmark of the collective proximity effect \cite{feigelman2008,eley2012,chen2024}, which persists even when removing most of the islands from a large graphene area [see Fig.~5 in SM \cite{SI}].

Our results show that the S/N interface conductance is a key parameter controlling the proximity effect in graphene, which depends on the degree of doping. In particular, electron doped graphene has a similar workfuncion as lead, resulting in smaller interface dipoles and smaller tunneling rates compared to neutral graphene \cite{Mammadov2017}. The graphene-metal separation also tends to increase with the degree of electron doping \cite{Giovannetti2008}, contributing to reducing the interface transparency. Based on this results, the uniformity of this superconducting state in graphene can be gate-controlled \cite{eley2012,han2014a}.

\begin{acknowledgments}
We acknowledge financial support from the Spanish MCIN/AEI/10.13039/501100011033 and the European Regional Development Fund (ERDF)  
through grants 
PID2022-140845OB-C61, 
PID2023-148225NB-C31, 
PID2023-149106NB-I00, 
CEX2020-001038-M,
CEX2023-001316-M, 
TED2021-130292B-C42, 
from the Basque Government through grant IT-1591-22, 
from the European Union NextGenerationEU/PRTR-C17.I1 through the IKUR Strategy of the Department of Education of the Basque Government under a collaboration agreement with Ikerbasque MPC, CIC nanoGUNE, and DIPC, 
the Comunidad de Madrid and the Spanish State through the Recovery, Transformation and Resilience Plan [Materiales Disruptivos Bidimensionales (2D), (MAD2DCM)-UAM Materiales Avanzados], 
and from the European Union’s Horizon Europe 
through grants ERC-AdG CONSPIRA (No. 101097693) 
and  JOSEPHINE (No. 101130224).  
J.O. acknowledges the scholarship PRE-2021-1-0350 from the Basque Government. BV acknowledges funding from the Spanish Ministerio de Universidades through the PhD scholarship No. FPU22/03675. E.C.R acknowledges funding from the Alexander von Humboldt Foundation via the Henriette Herz program. 
 
\end{acknowledgments}



\begin{thebibliography}{61}%
\makeatletter
\providecommand \@ifxundefined [1]{%
 \@ifx{#1\undefined}
}%
\providecommand \@ifnum [1]{%
 \ifnum #1\expandafter \@firstoftwo
 \else \expandafter \@secondoftwo
 \fi
}%
\providecommand \@ifx [1]{%
 \ifx #1\expandafter \@firstoftwo
 \else \expandafter \@secondoftwo
 \fi
}%
\providecommand \natexlab [1]{#1}%
\providecommand \enquote  [1]{``#1''}%
\providecommand \bibnamefont  [1]{#1}%
\providecommand \bibfnamefont [1]{#1}%
\providecommand \citenamefont [1]{#1}%
\providecommand \href@noop [0]{\@secondoftwo}%
\providecommand \href [0]{\begingroup \@sanitize@url \@href}%
\providecommand \@href[1]{\@@startlink{#1}\@@href}%
\providecommand \@@href[1]{\endgroup#1\@@endlink}%
\providecommand \@sanitize@url [0]{\catcode `\\12\catcode `\$12\catcode `\&12\catcode `\#12\catcode `\^12\catcode `\_12\catcode `\%12\relax}%
\providecommand \@@startlink[1]{}%
\providecommand \@@endlink[0]{}%
\providecommand \url  [0]{\begingroup\@sanitize@url \@url }%
\providecommand \@url [1]{\endgroup\@href {#1}{\urlprefix }}%
\providecommand \urlprefix  [0]{URL }%
\providecommand \Eprint [0]{\href }%
\providecommand \doibase [0]{http://dx.doi.org/}%
\providecommand \selectlanguage [0]{\@gobble}%
\providecommand \bibinfo  [0]{\@secondoftwo}%
\providecommand \bibfield  [0]{\@secondoftwo}%
\providecommand \translation [1]{[#1]}%
\providecommand \BibitemOpen [0]{}%
\providecommand \bibitemStop [0]{}%
\providecommand \bibitemNoStop [0]{.\EOS\space}%
\providecommand \EOS [0]{\spacefactor3000\relax}%
\providecommand \BibitemShut  [1]{\csname bibitem#1\endcsname}%
\let\auto@bib@innerbib\@empty
\bibitem [{\citenamefont {Andreev}(1964)}]{andreev1964}%
  \BibitemOpen
  \bibfield  {author} {\bibinfo {author} {\bibfnamefont {A.}~\bibnamefont {Andreev}},\ }\bibfield  {title} {\enquote {\bibinfo {title} {The {{Thermal Conductivity}} of the {{Intermediate State}} in {{Superconductors}}},}\ }\href@noop {} {\bibfield  {journal} {\bibinfo  {journal} {JETP Lett.}\ }\textbf {\bibinfo {volume} {19}},\ \bibinfo {pages} {1823.1828} (\bibinfo {year} {1964})}\BibitemShut {NoStop}%
\bibitem [{\citenamefont {Klapwijk}(2004)}]{klapwijk2004}%
  \BibitemOpen
  \bibfield  {author} {\bibinfo {author} {\bibfnamefont {T.~M.}\ \bibnamefont {Klapwijk}},\ }\bibfield  {title} {\enquote {\bibinfo {title} {Proximity {{Effect From}} an {{Andreev Perspective}}},}\ }\href {\doibase 10.1007/s10948-004-0773-0} {\bibfield  {journal} {\bibinfo  {journal} {J. Supercond.}\ }\textbf {\bibinfo {volume} {17}},\ \bibinfo {pages} {593--611} (\bibinfo {year} {2004})}\BibitemShut {NoStop}%
\bibitem [{\citenamefont {Bergeret}\ \emph {et~al.}(2005)\citenamefont {Bergeret}, \citenamefont {Volkov},\ and\ \citenamefont {Efetov}}]{bergeret2005}%
  \BibitemOpen
  \bibfield  {author} {\bibinfo {author} {\bibfnamefont {F.~S.}\ \bibnamefont {Bergeret}}, \bibinfo {author} {\bibfnamefont {A.~F.}\ \bibnamefont {Volkov}}, \ and\ \bibinfo {author} {\bibfnamefont {K.~B.}\ \bibnamefont {Efetov}},\ }\bibfield  {title} {\enquote {\bibinfo {title} {Odd triplet superconductivity and related phenomena in superconductor-ferromagnet structures},}\ }\href {\doibase 10.1103/RevModPhys.77.1321} {\bibfield  {journal} {\bibinfo  {journal} {Rev. Mod. Phys.}\ }\textbf {\bibinfo {volume} {77}},\ \bibinfo {pages} {1321--1373} (\bibinfo {year} {2005})}\BibitemShut {NoStop}%
\bibitem [{\citenamefont {{Black-Schaffer}}\ and\ \citenamefont {Balatsky}(2013)}]{black-schaffer2013}%
  \BibitemOpen
  \bibfield  {author} {\bibinfo {author} {\bibfnamefont {A.~M.}\ \bibnamefont {{Black-Schaffer}}}\ and\ \bibinfo {author} {\bibfnamefont {A.~V.}\ \bibnamefont {Balatsky}},\ }\bibfield  {title} {\enquote {\bibinfo {title} {Proximity-induced unconventional superconductivity in topological insulators},}\ }\href {\doibase 10.1103/PhysRevB.87.220506} {\bibfield  {journal} {\bibinfo  {journal} {Phys. Rev. B}\ }\textbf {\bibinfo {volume} {87}},\ \bibinfo {pages} {220506} (\bibinfo {year} {2013})}\BibitemShut {NoStop}%
\bibitem [{\citenamefont {Cao}\ \emph {et~al.}(2018)\citenamefont {Cao}, \citenamefont {Fatemi}, \citenamefont {Fang}, \citenamefont {Watanabe}, \citenamefont {Taniguchi}, \citenamefont {Kaxiras},\ and\ \citenamefont {{Jarillo-Herrero}}}]{cao2018}%
  \BibitemOpen
  \bibfield  {author} {\bibinfo {author} {\bibfnamefont {Y.}~\bibnamefont {Cao}}, \bibinfo {author} {\bibfnamefont {V.}~\bibnamefont {Fatemi}}, \bibinfo {author} {\bibfnamefont {S.}~\bibnamefont {Fang}}, \bibinfo {author} {\bibfnamefont {K.}~\bibnamefont {Watanabe}}, \bibinfo {author} {\bibfnamefont {T.}~\bibnamefont {Taniguchi}}, \bibinfo {author} {\bibfnamefont {E.}~\bibnamefont {Kaxiras}}, \ and\ \bibinfo {author} {\bibfnamefont {P.}~\bibnamefont {{Jarillo-Herrero}}},\ }\bibfield  {title} {\enquote {\bibinfo {title} {Unconventional superconductivity in magic-angle graphene superlattices},}\ }\href {\doibase 10.1038/nature26160} {\bibfield  {journal} {\bibinfo  {journal} {Nature}\ }\textbf {\bibinfo {volume} {556}},\ \bibinfo {pages} {43--50} (\bibinfo {year} {2018})}\BibitemShut {NoStop}%
\bibitem [{\citenamefont {Balents}\ \emph {et~al.}(2020)\citenamefont {Balents}, \citenamefont {Dean}, \citenamefont {Efetov},\ and\ \citenamefont {Young}}]{balents2020}%
  \BibitemOpen
  \bibfield  {author} {\bibinfo {author} {\bibfnamefont {L.}~\bibnamefont {Balents}}, \bibinfo {author} {\bibfnamefont {C.~R.}\ \bibnamefont {Dean}}, \bibinfo {author} {\bibfnamefont {D.~K.}\ \bibnamefont {Efetov}}, \ and\ \bibinfo {author} {\bibfnamefont {A.~F.}\ \bibnamefont {Young}},\ }\bibfield  {title} {\enquote {\bibinfo {title} {Superconductivity and strong correlations in moir{\'e} flat bands},}\ }\href {\doibase 10.1038/s41567-020-0906-9} {\bibfield  {journal} {\bibinfo  {journal} {Nat. Phys.}\ }\textbf {\bibinfo {volume} {16}},\ \bibinfo {pages} {725--733} (\bibinfo {year} {2020})}\BibitemShut {NoStop}%
\bibitem [{\citenamefont {Leijnse}\ and\ \citenamefont {Flensberg}(2012)}]{leijnse2012}%
  \BibitemOpen
  \bibfield  {author} {\bibinfo {author} {\bibfnamefont {M.}~\bibnamefont {Leijnse}}\ and\ \bibinfo {author} {\bibfnamefont {K.}~\bibnamefont {Flensberg}},\ }\bibfield  {title} {\enquote {\bibinfo {title} {Introduction to topological superconductivity and {{Majorana}} fermions},}\ }\href {\doibase 10.1088/0268-1242/27/12/124003} {\bibfield  {journal} {\bibinfo  {journal} {Semiconductor Science and Technology}\ }\textbf {\bibinfo {volume} {27}},\ \bibinfo {pages} {124003} (\bibinfo {year} {2012})}\BibitemShut {NoStop}%
\bibitem [{\citenamefont {{San-Jose}}\ \emph {et~al.}(2015)\citenamefont {{San-Jose}}, \citenamefont {Lado}, \citenamefont {Aguado}, \citenamefont {Guinea},\ and\ \citenamefont {{Fern{\'a}ndez-Rossier}}}]{san-jose2015}%
  \BibitemOpen
  \bibfield  {author} {\bibinfo {author} {\bibfnamefont {P.}~\bibnamefont {{San-Jose}}}, \bibinfo {author} {\bibfnamefont {J.~L.}\ \bibnamefont {Lado}}, \bibinfo {author} {\bibfnamefont {R.}~\bibnamefont {Aguado}}, \bibinfo {author} {\bibfnamefont {F.}~\bibnamefont {Guinea}}, \ and\ \bibinfo {author} {\bibfnamefont {J.}~\bibnamefont {{Fern{\'a}ndez-Rossier}}},\ }\bibfield  {title} {\enquote {\bibinfo {title} {Majorana {{Zero Modes}} in {{Graphene}}},}\ }\href {\doibase 10.1103/PhysRevX.5.041042} {\bibfield  {journal} {\bibinfo  {journal} {Phys. Rev. X}\ }\textbf {\bibinfo {volume} {5}},\ \bibinfo {pages} {041042} (\bibinfo {year} {2015})}\BibitemShut {NoStop}%
\bibitem [{\citenamefont {Frolov}\ \emph {et~al.}(2020)\citenamefont {Frolov}, \citenamefont {Manfra},\ and\ \citenamefont {Sau}}]{frolov2020}%
  \BibitemOpen
  \bibfield  {author} {\bibinfo {author} {\bibfnamefont {S.~M.}\ \bibnamefont {Frolov}}, \bibinfo {author} {\bibfnamefont {M.~J.}\ \bibnamefont {Manfra}}, \ and\ \bibinfo {author} {\bibfnamefont {J.~D.}\ \bibnamefont {Sau}},\ }\bibfield  {title} {\enquote {\bibinfo {title} {Topological superconductivity in hybrid devices},}\ }\href {\doibase 10.1038/s41567-020-0925-6} {\bibfield  {journal} {\bibinfo  {journal} {Nature Physics}\ }\textbf {\bibinfo {volume} {16}},\ \bibinfo {pages} {718--724} (\bibinfo {year} {2020})}\BibitemShut {NoStop}%
\bibitem [{\citenamefont {Ando}\ \emph {et~al.}(2020)\citenamefont {Ando}, \citenamefont {Miyasaka}, \citenamefont {Li}, \citenamefont {Ishizuka}, \citenamefont {Arakawa}, \citenamefont {Shiota}, \citenamefont {Moriyama}, \citenamefont {Yanase},\ and\ \citenamefont {Ono}}]{ando2020}%
  \BibitemOpen
  \bibfield  {author} {\bibinfo {author} {\bibfnamefont {F.}~\bibnamefont {Ando}}, \bibinfo {author} {\bibfnamefont {Y.}~\bibnamefont {Miyasaka}}, \bibinfo {author} {\bibfnamefont {T.}~\bibnamefont {Li}}, \bibinfo {author} {\bibfnamefont {J.}~\bibnamefont {Ishizuka}}, \bibinfo {author} {\bibfnamefont {T.}~\bibnamefont {Arakawa}}, \bibinfo {author} {\bibfnamefont {Y.}~\bibnamefont {Shiota}}, \bibinfo {author} {\bibfnamefont {T.}~\bibnamefont {Moriyama}}, \bibinfo {author} {\bibfnamefont {Y.}~\bibnamefont {Yanase}}, \ and\ \bibinfo {author} {\bibfnamefont {T.}~\bibnamefont {Ono}},\ }\bibfield  {title} {\enquote {\bibinfo {title} {Observation of superconducting diode effect},}\ }\href {\doibase 10.1038/s41586-020-2590-4} {\bibfield  {journal} {\bibinfo  {journal} {Nature}\ }\textbf {\bibinfo {volume} {584}},\ \bibinfo {pages} {373--376} (\bibinfo {year} {2020})}\BibitemShut {NoStop}%
\bibitem [{\citenamefont {{Ili{\'c} {\'c}}}\ \emph {et~al.}(2024)\citenamefont {{Ili{\'c} {\'c}}}, \citenamefont {Virtanen}, \citenamefont {Crawford}, \citenamefont {Heikkil{\"a}},\ and\ \citenamefont {Bergeret}}]{ilicc2024}%
  \BibitemOpen
  \bibfield  {author} {\bibinfo {author} {\bibfnamefont {S.}~\bibnamefont {{Ili{\'c} {\'c}}}}, \bibinfo {author} {\bibfnamefont {P.}~\bibnamefont {Virtanen}}, \bibinfo {author} {\bibfnamefont {D.}~\bibnamefont {Crawford}}, \bibinfo {author} {\bibfnamefont {T.~T.}\ \bibnamefont {Heikkil{\"a}}}, \ and\ \bibinfo {author} {\bibfnamefont {F.~S.}\ \bibnamefont {Bergeret}},\ }\bibfield  {title} {\enquote {\bibinfo {title} {Superconducting diode effect in diffusive superconductors and {{Josephson}} junctions with {{Rashba}} spin-orbit coupling},}\ }\href {\doibase 10.1103/PhysRevB.110.L140501} {\bibfield  {journal} {\bibinfo  {journal} {Phys. Rev. B}\ }\textbf {\bibinfo {volume} {110}},\ \bibinfo {pages} {L140501} (\bibinfo {year} {2024})}\BibitemShut {NoStop}%
\bibitem [{\citenamefont {Kezilebieke}\ \emph {et~al.}(2020)\citenamefont {Kezilebieke}, \citenamefont {Huda}, \citenamefont {Va{\v n}o}, \citenamefont {Aapro}, \citenamefont {Ganguli}, \citenamefont {Silveira}, \citenamefont {G{\l}odzik}, \citenamefont {Foster}, \citenamefont {Ojanen},\ and\ \citenamefont {Liljeroth}}]{kezilebieke2020b}%
  \BibitemOpen
  \bibfield  {author} {\bibinfo {author} {\bibfnamefont {S.}~\bibnamefont {Kezilebieke}}, \bibinfo {author} {\bibfnamefont {M.~N.}\ \bibnamefont {Huda}}, \bibinfo {author} {\bibfnamefont {V.}~\bibnamefont {Va{\v n}o}}, \bibinfo {author} {\bibfnamefont {M.}~\bibnamefont {Aapro}}, \bibinfo {author} {\bibfnamefont {S.~C.}\ \bibnamefont {Ganguli}}, \bibinfo {author} {\bibfnamefont {O.~J.}\ \bibnamefont {Silveira}}, \bibinfo {author} {\bibfnamefont {S.}~\bibnamefont {G{\l}odzik}}, \bibinfo {author} {\bibfnamefont {A.~S.}\ \bibnamefont {Foster}}, \bibinfo {author} {\bibfnamefont {T.}~\bibnamefont {Ojanen}}, \ and\ \bibinfo {author} {\bibfnamefont {P.}~\bibnamefont {Liljeroth}},\ }\bibfield  {title} {\enquote {\bibinfo {title} {Topological superconductivity in a van der {{Waals}} heterostructure},}\ }\href {\doibase 10.1038/s41586-020-2989-y} {\bibfield  {journal} {\bibinfo  {journal} {Nature}\ }\textbf {\bibinfo {volume} {588}},\ \bibinfo {pages} {424--428} (\bibinfo {year} {2020})}\BibitemShut {NoStop}%
\bibitem [{\citenamefont {Manna}\ \emph {et~al.}(2020)\citenamefont {Manna}, \citenamefont {Wei}, \citenamefont {Xie}, \citenamefont {Law}, \citenamefont {Lee},\ and\ \citenamefont {Moodera}}]{manna2020a}%
  \BibitemOpen
  \bibfield  {author} {\bibinfo {author} {\bibfnamefont {S.}~\bibnamefont {Manna}}, \bibinfo {author} {\bibfnamefont {P.}~\bibnamefont {Wei}}, \bibinfo {author} {\bibfnamefont {Y.}~\bibnamefont {Xie}}, \bibinfo {author} {\bibfnamefont {K.~T.}\ \bibnamefont {Law}}, \bibinfo {author} {\bibfnamefont {P.~A.}\ \bibnamefont {Lee}}, \ and\ \bibinfo {author} {\bibfnamefont {J.~S.}\ \bibnamefont {Moodera}},\ }\bibfield  {title} {\enquote {\bibinfo {title} {Signature of a pair of {{Majorana}} zero modes in superconducting gold surface states},}\ }\href {\doibase 10.1073/pnas.1919753117} {\bibfield  {journal} {\bibinfo  {journal} {Proc. Natl. Acad. Sci. U.S.A.}\ }\textbf {\bibinfo {volume} {117}},\ \bibinfo {pages} {8775--8782} (\bibinfo {year} {2020})}\BibitemShut {NoStop}%
\bibitem [{\citenamefont {Stolyarov}\ \emph {et~al.}(2021)\citenamefont {Stolyarov}, \citenamefont {Pons}, \citenamefont {Vlaic}, \citenamefont {Remizov}, \citenamefont {Shapiro}, \citenamefont {Brun}, \citenamefont {Bozhko}, \citenamefont {Cren}, \citenamefont {Menshchikova}, \citenamefont {Chulkov}, \citenamefont {Pogosov}, \citenamefont {Lozovik},\ and\ \citenamefont {Roditchev}}]{stolyarov2021}%
  \BibitemOpen
  \bibfield  {author} {\bibinfo {author} {\bibfnamefont {V.~S.}\ \bibnamefont {Stolyarov}}, \bibinfo {author} {\bibfnamefont {S.}~\bibnamefont {Pons}}, \bibinfo {author} {\bibfnamefont {S.}~\bibnamefont {Vlaic}}, \bibinfo {author} {\bibfnamefont {S.~V.}\ \bibnamefont {Remizov}}, \bibinfo {author} {\bibfnamefont {D.~S.}\ \bibnamefont {Shapiro}}, \bibinfo {author} {\bibfnamefont {C.}~\bibnamefont {Brun}}, \bibinfo {author} {\bibfnamefont {S.~I.}\ \bibnamefont {Bozhko}}, \bibinfo {author} {\bibfnamefont {T.}~\bibnamefont {Cren}}, \bibinfo {author} {\bibfnamefont {T.~V.}\ \bibnamefont {Menshchikova}}, \bibinfo {author} {\bibfnamefont {E.~V.}\ \bibnamefont {Chulkov}}, \bibinfo {author} {\bibfnamefont {W.~V.}\ \bibnamefont {Pogosov}}, \bibinfo {author} {\bibfnamefont {Y.~E.}\ \bibnamefont {Lozovik}}, \ and\ \bibinfo {author} {\bibfnamefont {D.}~\bibnamefont {Roditchev}},\ }\bibfield  {title} {\enquote {\bibinfo {title} {Superconducting {{Long-Range Proximity Effect}} through the {{Atomically Flat Interface}} of a
  {{Bi}} {\textsubscript{2}} {{Te}} {\textsubscript{3}} {{Topological Insulator}}},}\ }\href {\doibase 10.1021/acs.jpclett.1c02257} {\bibfield  {journal} {\bibinfo  {journal} {J. Phys. Chem. Lett.}\ }\textbf {\bibinfo {volume} {12}},\ \bibinfo {pages} {9068--9075} (\bibinfo {year} {2021})}\BibitemShut {NoStop}%
\bibitem [{\citenamefont {Vaxevani}\ \emph {et~al.}(2022)\citenamefont {Vaxevani}, \citenamefont {Li}, \citenamefont {Trivini}, \citenamefont {Ortuzar}, \citenamefont {Longo}, \citenamefont {Wang},\ and\ \citenamefont {Pascual}}]{vaxevani2022}%
  \BibitemOpen
  \bibfield  {author} {\bibinfo {author} {\bibfnamefont {K.}~\bibnamefont {Vaxevani}}, \bibinfo {author} {\bibfnamefont {J.}~\bibnamefont {Li}}, \bibinfo {author} {\bibfnamefont {S.}~\bibnamefont {Trivini}}, \bibinfo {author} {\bibfnamefont {J.}~\bibnamefont {Ortuzar}}, \bibinfo {author} {\bibfnamefont {D.}~\bibnamefont {Longo}}, \bibinfo {author} {\bibfnamefont {D.}~\bibnamefont {Wang}}, \ and\ \bibinfo {author} {\bibfnamefont {J.~I.}\ \bibnamefont {Pascual}},\ }\bibfield  {title} {\enquote {\bibinfo {title} {Extending the {{Spin Excitation Lifetime}} of a {{Magnetic Molecule}} on a {{Proximitized Superconductor}}},}\ }\href {\doibase 10.1021/acs.nanolett.2c00924} {\bibfield  {journal} {\bibinfo  {journal} {Nano Lett.}\ }\textbf {\bibinfo {volume} {22}},\ \bibinfo {pages} {6075--6082} (\bibinfo {year} {2022})},\ \bibinfo {note} {doi: 10.1021/acs.nanolett.2c00924}\BibitemShut {NoStop}%
\bibitem [{\citenamefont {Trivini}\ \emph {et~al.}(2023)\citenamefont {Trivini}, \citenamefont {Ortuzar}, \citenamefont {Vaxevani}, \citenamefont {Li}, \citenamefont {Bergeret}, \citenamefont {Cazalilla},\ and\ \citenamefont {Pascual}}]{trivini2023a}%
  \BibitemOpen
  \bibfield  {author} {\bibinfo {author} {\bibfnamefont {S.}~\bibnamefont {Trivini}}, \bibinfo {author} {\bibfnamefont {J.}~\bibnamefont {Ortuzar}}, \bibinfo {author} {\bibfnamefont {K.}~\bibnamefont {Vaxevani}}, \bibinfo {author} {\bibfnamefont {J.}~\bibnamefont {Li}}, \bibinfo {author} {\bibfnamefont {F.~S.}\ \bibnamefont {Bergeret}}, \bibinfo {author} {\bibfnamefont {M.~A.}\ \bibnamefont {Cazalilla}}, \ and\ \bibinfo {author} {\bibfnamefont {J.~I.}\ \bibnamefont {Pascual}},\ }\bibfield  {title} {\enquote {\bibinfo {title} {Cooper {{Pair Excitation Mediated}} by a {{Molecular Quantum Spin}} on a {{Superconducting Proximitized Gold Film}}},}\ }\href {\doibase 10.1103/PhysRevLett.130.136004} {\bibfield  {journal} {\bibinfo  {journal} {Phys. Rev. Lett.}\ }\textbf {\bibinfo {volume} {130}},\ \bibinfo {pages} {136004} (\bibinfo {year} {2023})}\BibitemShut {NoStop}%
\bibitem [{\citenamefont {Usadel}(1970)}]{usadel1970}%
  \BibitemOpen
  \bibfield  {author} {\bibinfo {author} {\bibfnamefont {K.~D.}\ \bibnamefont {Usadel}},\ }\bibfield  {title} {\enquote {\bibinfo {title} {Generalized {{Diffusion Equation}} for {{Superconducting Alloys}}},}\ }\href {\doibase 10.1103/PhysRevLett.25.507} {\bibfield  {journal} {\bibinfo  {journal} {Phys. Rev. Lett.}\ }\textbf {\bibinfo {volume} {25}},\ \bibinfo {pages} {507--509} (\bibinfo {year} {1970})}\BibitemShut {NoStop}%
\bibitem [{\citenamefont {Fominov}\ and\ \citenamefont {Feigel'man}(2001)}]{fominov2001superconductive}%
  \BibitemOpen
  \bibfield  {author} {\bibinfo {author} {\bibfnamefont {{\relax Ya}.~V.}\ \bibnamefont {Fominov}}\ and\ \bibinfo {author} {\bibfnamefont {M.~V.}\ \bibnamefont {Feigel'man}},\ }\bibfield  {title} {\enquote {\bibinfo {title} {Superconductive properties of thin dirty superconductor--normal-metal bilayers},}\ }\href {\doibase 10.1103/PhysRevB.63.094518} {\bibfield  {journal} {\bibinfo  {journal} {Phys. Rev. B}\ }\textbf {\bibinfo {volume} {63}},\ \bibinfo {pages} {094518} (\bibinfo {year} {2001})}\BibitemShut {NoStop}%
\bibitem [{\citenamefont {Golubov}\ \emph {et~al.}(2004)\citenamefont {Golubov}, \citenamefont {Kupriyanov},\ and\ \citenamefont {Il'ichev}}]{golubov2004the}%
  \BibitemOpen
  \bibfield  {author} {\bibinfo {author} {\bibfnamefont {A.~A.}\ \bibnamefont {Golubov}}, \bibinfo {author} {\bibfnamefont {M.~{\relax Yu}.}\ \bibnamefont {Kupriyanov}}, \ and\ \bibinfo {author} {\bibfnamefont {E.}~\bibnamefont {Il'ichev}},\ }\bibfield  {title} {\enquote {\bibinfo {title} {The current-phase relation in {{Josephson}} junctions},}\ }\href {\doibase 10.1103/RevModPhys.76.411} {\bibfield  {journal} {\bibinfo  {journal} {Rev. Mod. Phys.}\ }\textbf {\bibinfo {volume} {76}},\ \bibinfo {pages} {411--469} (\bibinfo {year} {2004})}\BibitemShut {NoStop}%
\bibitem [{\citenamefont {Giazotto}\ \emph {et~al.}(2010)\citenamefont {Giazotto}, \citenamefont {Peltonen}, \citenamefont {Meschke},\ and\ \citenamefont {Pekola}}]{giazotto2010superconducting}%
  \BibitemOpen
  \bibfield  {author} {\bibinfo {author} {\bibfnamefont {F.}~\bibnamefont {Giazotto}}, \bibinfo {author} {\bibfnamefont {J.~T.}\ \bibnamefont {Peltonen}}, \bibinfo {author} {\bibfnamefont {M.}~\bibnamefont {Meschke}}, \ and\ \bibinfo {author} {\bibfnamefont {J.~P.}\ \bibnamefont {Pekola}},\ }\bibfield  {title} {\enquote {\bibinfo {title} {Superconducting quantum interference proximity transistor},}\ }\href {\doibase 10.1038/nphys1537} {\bibfield  {journal} {\bibinfo  {journal} {Nature Physics}\ }\textbf {\bibinfo {volume} {6}},\ \bibinfo {pages} {254--259} (\bibinfo {year} {2010})}\BibitemShut {NoStop}%
\bibitem [{\citenamefont {Cherkez}\ \emph {et~al.}(2014)\citenamefont {Cherkez}, \citenamefont {Cuevas}, \citenamefont {Brun}, \citenamefont {Cren}, \citenamefont {M{\'e}nard}, \citenamefont {Debontridder}, \citenamefont {Stolyarov},\ and\ \citenamefont {Roditchev}}]{cherkez2014proximity}%
  \BibitemOpen
  \bibfield  {author} {\bibinfo {author} {\bibfnamefont {V.}~\bibnamefont {Cherkez}}, \bibinfo {author} {\bibfnamefont {J.~C.}\ \bibnamefont {Cuevas}}, \bibinfo {author} {\bibfnamefont {C.}~\bibnamefont {Brun}}, \bibinfo {author} {\bibfnamefont {T.}~\bibnamefont {Cren}}, \bibinfo {author} {\bibfnamefont {G.}~\bibnamefont {M{\'e}nard}}, \bibinfo {author} {\bibfnamefont {F.}~\bibnamefont {Debontridder}}, \bibinfo {author} {\bibfnamefont {V.~S.}\ \bibnamefont {Stolyarov}}, \ and\ \bibinfo {author} {\bibfnamefont {D.}~\bibnamefont {Roditchev}},\ }\bibfield  {title} {\enquote {\bibinfo {title} {Proximity effect between two superconductors spatially resolved by scanning tunneling spectroscopy},}\ }\href {\doibase 10.1103/PhysRevX.4.011033} {\bibfield  {journal} {\bibinfo  {journal} {Phys. Rev. X}\ }\textbf {\bibinfo {volume} {4}},\ \bibinfo {pages} {011033} (\bibinfo {year} {2014})}\BibitemShut {NoStop}%
\bibitem [{\citenamefont {Hijano}\ \emph {et~al.}(2021)\citenamefont {Hijano}, \citenamefont {{Ili{\'c} {\'c}}}, \citenamefont {Rouco}, \citenamefont {{Gonz{\'a}lez-Orellana}}, \citenamefont {Ilyn}, \citenamefont {Rogero}, \citenamefont {Virtanen}, \citenamefont {Heikkil{\"a}}, \citenamefont {Khorshidian}, \citenamefont {Spies}, \citenamefont {Ligato}, \citenamefont {Giazotto}, \citenamefont {Strambini},\ and\ \citenamefont {Bergeret}}]{hijano2021coexistence}%
  \BibitemOpen
  \bibfield  {author} {\bibinfo {author} {\bibfnamefont {A.}~\bibnamefont {Hijano}}, \bibinfo {author} {\bibfnamefont {S.}~\bibnamefont {{Ili{\'c} {\'c}}}}, \bibinfo {author} {\bibfnamefont {M.}~\bibnamefont {Rouco}}, \bibinfo {author} {\bibfnamefont {C.}~\bibnamefont {{Gonz{\'a}lez-Orellana}}}, \bibinfo {author} {\bibfnamefont {M.}~\bibnamefont {Ilyn}}, \bibinfo {author} {\bibfnamefont {C.}~\bibnamefont {Rogero}}, \bibinfo {author} {\bibfnamefont {P.}~\bibnamefont {Virtanen}}, \bibinfo {author} {\bibfnamefont {T.~T.}\ \bibnamefont {Heikkil{\"a}}}, \bibinfo {author} {\bibfnamefont {S.}~\bibnamefont {Khorshidian}}, \bibinfo {author} {\bibfnamefont {M.}~\bibnamefont {Spies}}, \bibinfo {author} {\bibfnamefont {N.}~\bibnamefont {Ligato}}, \bibinfo {author} {\bibfnamefont {F.}~\bibnamefont {Giazotto}}, \bibinfo {author} {\bibfnamefont {E.}~\bibnamefont {Strambini}}, \ and\ \bibinfo {author} {\bibfnamefont {F.~S.}\ \bibnamefont {Bergeret}},\ }\bibfield  {title} {\enquote {\bibinfo {title} {Coexistence of
  superconductivity and spin-splitting fields in superconductor/ferromagnetic insulator bilayers of arbitrary thickness},}\ }\href {\doibase 10.1103/PhysRevResearch.3.023131} {\bibfield  {journal} {\bibinfo  {journal} {Phys. Rev. Res.}\ }\textbf {\bibinfo {volume} {3}},\ \bibinfo {pages} {023131} (\bibinfo {year} {2021})}\BibitemShut {NoStop}%
\bibitem [{\citenamefont {Strambini}\ \emph {et~al.}(2017)\citenamefont {Strambini}, \citenamefont {Golovach}, \citenamefont {De~Simoni}, \citenamefont {Moodera}, \citenamefont {Bergeret},\ and\ \citenamefont {Giazotto}}]{strambini2017revealing}%
  \BibitemOpen
  \bibfield  {author} {\bibinfo {author} {\bibfnamefont {E.}~\bibnamefont {Strambini}}, \bibinfo {author} {\bibfnamefont {V.~N.}\ \bibnamefont {Golovach}}, \bibinfo {author} {\bibfnamefont {G.}~\bibnamefont {De~Simoni}}, \bibinfo {author} {\bibfnamefont {J.~S.}\ \bibnamefont {Moodera}}, \bibinfo {author} {\bibfnamefont {F.~S.}\ \bibnamefont {Bergeret}}, \ and\ \bibinfo {author} {\bibfnamefont {F.}~\bibnamefont {Giazotto}},\ }\bibfield  {title} {\enquote {\bibinfo {title} {Revealing the magnetic proximity effect in {{EuS}}/{{Al}} bilayers through superconducting tunneling spectroscopy},}\ }\href {\doibase 10.1103/PhysRevMaterials.1.054402} {\bibfield  {journal} {\bibinfo  {journal} {Phys. Rev. Mater.}\ }\textbf {\bibinfo {volume} {1}},\ \bibinfo {pages} {054402} (\bibinfo {year} {2017})}\BibitemShut {NoStop}%
\bibitem [{\citenamefont {Kashiwaya}\ \emph {et~al.}(1998)\citenamefont {Kashiwaya}, \citenamefont {Ito}, \citenamefont {Oka}, \citenamefont {Ueno}, \citenamefont {Takashima}, \citenamefont {Koyanagi}, \citenamefont {Tanaka},\ and\ \citenamefont {Kajimura}}]{kashiwaya1998tunneling}%
  \BibitemOpen
  \bibfield  {author} {\bibinfo {author} {\bibfnamefont {S.}~\bibnamefont {Kashiwaya}}, \bibinfo {author} {\bibfnamefont {T.}~\bibnamefont {Ito}}, \bibinfo {author} {\bibfnamefont {K.}~\bibnamefont {Oka}}, \bibinfo {author} {\bibfnamefont {S.}~\bibnamefont {Ueno}}, \bibinfo {author} {\bibfnamefont {H.}~\bibnamefont {Takashima}}, \bibinfo {author} {\bibfnamefont {M.}~\bibnamefont {Koyanagi}}, \bibinfo {author} {\bibfnamefont {Y.}~\bibnamefont {Tanaka}}, \ and\ \bibinfo {author} {\bibfnamefont {K.}~\bibnamefont {Kajimura}},\ }\bibfield  {title} {\enquote {\bibinfo {title} {Tunneling spectroscopy of superconducting {{Nd}}{\textsubscript{1.85}}{{Ce}}{\textsubscript{0.15}}{{CuO}}{\textsubscript{4-{$\delta$}}}},}\ }\href {\doibase 10.1103/PhysRevB.57.8680} {\bibfield  {journal} {\bibinfo  {journal} {Phys. Rev. B}\ }\textbf {\bibinfo {volume} {57}},\ \bibinfo {pages} {8680--8686} (\bibinfo {year} {1998})}\BibitemShut {NoStop}%
\bibitem [{\citenamefont {Kashiwaya}\ \emph {et~al.}(1995)\citenamefont {Kashiwaya}, \citenamefont {Tanaka}, \citenamefont {Koyanagi}, \citenamefont {Takashima},\ and\ \citenamefont {Kajimura}}]{kashiwaya1995origin}%
  \BibitemOpen
  \bibfield  {author} {\bibinfo {author} {\bibfnamefont {S.}~\bibnamefont {Kashiwaya}}, \bibinfo {author} {\bibfnamefont {Y.}~\bibnamefont {Tanaka}}, \bibinfo {author} {\bibfnamefont {M.}~\bibnamefont {Koyanagi}}, \bibinfo {author} {\bibfnamefont {H.}~\bibnamefont {Takashima}}, \ and\ \bibinfo {author} {\bibfnamefont {K.}~\bibnamefont {Kajimura}},\ }\bibfield  {title} {\enquote {\bibinfo {title} {Origin of zero-bias conductance peaks in high-{{T}}{\textsubscript{c}} superconductors},}\ }\href {\doibase 10.1103/PhysRevB.51.1350} {\bibfield  {journal} {\bibinfo  {journal} {Phys. Rev. B}\ }\textbf {\bibinfo {volume} {51}},\ \bibinfo {pages} {1350--1353} (\bibinfo {year} {1995})}\BibitemShut {NoStop}%
\bibitem [{\citenamefont {Horcas}\ \emph {et~al.}(2007)\citenamefont {Horcas}, \citenamefont {Fern{\'a}ndez}, \citenamefont {{G{\'o}mez-Rodr{\'i}guez}}, \citenamefont {Colchero}, \citenamefont {{G{\'o}mez-Herrero}},\ and\ \citenamefont {Baro}}]{horcas2007}%
  \BibitemOpen
  \bibfield  {author} {\bibinfo {author} {\bibfnamefont {I.}~\bibnamefont {Horcas}}, \bibinfo {author} {\bibfnamefont {R.}~\bibnamefont {Fern{\'a}ndez}}, \bibinfo {author} {\bibfnamefont {J.~M.}\ \bibnamefont {{G{\'o}mez-Rodr{\'i}guez}}}, \bibinfo {author} {\bibfnamefont {J.}~\bibnamefont {Colchero}}, \bibinfo {author} {\bibfnamefont {J.}~\bibnamefont {{G{\'o}mez-Herrero}}}, \ and\ \bibinfo {author} {\bibfnamefont {A.~M.}\ \bibnamefont {Baro}},\ }\bibfield  {title} {\enquote {\bibinfo {title} {{{WSXM}}: {{A}} software for scanning probe microscopy and a tool for nanotechnology},}\ }\href {\doibase 10.1063/1.2432410} {\bibfield  {journal} {\bibinfo  {journal} {Rev. Sci. Inst.}\ }\textbf {\bibinfo {volume} {78}},\ \bibinfo {pages} {013705} (\bibinfo {year} {2007})}\BibitemShut {NoStop}%
\bibitem [{\citenamefont {Blonder}\ \emph {et~al.}(1982)\citenamefont {Blonder}, \citenamefont {Tinkham},\ and\ \citenamefont {Klapwijk}}]{blonder1982a}%
  \BibitemOpen
  \bibfield  {author} {\bibinfo {author} {\bibfnamefont {G.~E.}\ \bibnamefont {Blonder}}, \bibinfo {author} {\bibfnamefont {M.}~\bibnamefont {Tinkham}}, \ and\ \bibinfo {author} {\bibfnamefont {T.~M.}\ \bibnamefont {Klapwijk}},\ }\bibfield  {title} {\enquote {\bibinfo {title} {Transition from metallic to tunneling regimes in superconducting microconstrictions: {{Excess}} current, charge imbalance, and supercurrent conversion},}\ }\href {\doibase 10.1103/PhysRevB.25.4515} {\bibfield  {journal} {\bibinfo  {journal} {Phys. Rev. B}\ }\textbf {\bibinfo {volume} {25}},\ \bibinfo {pages} {4515--4532} (\bibinfo {year} {1982})}\BibitemShut {NoStop}%
\bibitem [{\citenamefont {Kuprianov}\ and\ \citenamefont {Lukichev}(1988)}]{kuprianov1988}%
  \BibitemOpen
  \bibfield  {author} {\bibinfo {author} {\bibfnamefont {M.~Y.}\ \bibnamefont {Kuprianov}}\ and\ \bibinfo {author} {\bibfnamefont {{\relax VF}.}~\bibnamefont {Lukichev}},\ }\bibfield  {title} {\enquote {\bibinfo {title} {Influence of boundary transparency on the critical current of ``dirty'' {{SS}}'{{S}} structures},}\ }\href {\doibase http://jetp.ras.ru/cgi-bin/e/index/e/67/6/p1163?a=list} {\bibfield  {journal} {\bibinfo  {journal} {Zh. Eksp. Teor. Fiz}\ }\textbf {\bibinfo {volume} {94}},\ \bibinfo {pages} {139} (\bibinfo {year} {1988})}\BibitemShut {NoStop}%
\bibitem [{\citenamefont {Belzig}\ \emph {et~al.}(1996)\citenamefont {Belzig}, \citenamefont {Bruder},\ and\ \citenamefont {Sch{\"o}n}}]{belzig1996}%
  \BibitemOpen
  \bibfield  {author} {\bibinfo {author} {\bibfnamefont {W.}~\bibnamefont {Belzig}}, \bibinfo {author} {\bibfnamefont {C.}~\bibnamefont {Bruder}}, \ and\ \bibinfo {author} {\bibfnamefont {G.}~\bibnamefont {Sch{\"o}n}},\ }\bibfield  {title} {\enquote {\bibinfo {title} {Local density of states in a dirty normal metal connected to a superconductor},}\ }\href {\doibase 10.1103/PhysRevB.54.9443} {\bibfield  {journal} {\bibinfo  {journal} {Phys. Rev. B}\ }\textbf {\bibinfo {volume} {54}},\ \bibinfo {pages} {9443--9448} (\bibinfo {year} {1996})}\BibitemShut {NoStop}%
\bibitem [{\citenamefont {Hammer}\ \emph {et~al.}(2007)\citenamefont {Hammer}, \citenamefont {Cuevas}, \citenamefont {Bergeret},\ and\ \citenamefont {Belzig}}]{hammer2007}%
  \BibitemOpen
  \bibfield  {author} {\bibinfo {author} {\bibfnamefont {J.~C.}\ \bibnamefont {Hammer}}, \bibinfo {author} {\bibfnamefont {J.~C.}\ \bibnamefont {Cuevas}}, \bibinfo {author} {\bibfnamefont {F.~S.}\ \bibnamefont {Bergeret}}, \ and\ \bibinfo {author} {\bibfnamefont {W.}~\bibnamefont {Belzig}},\ }\bibfield  {title} {\enquote {\bibinfo {title} {Density of states and supercurrent in diffusive {{SNS}} junctions: {{Roles}} of nonideal interfaces and spin-flip scattering},}\ }\href {\doibase 10.1103/PhysRevB.76.064514} {\bibfield  {journal} {\bibinfo  {journal} {Phys. Rev. B}\ }\textbf {\bibinfo {volume} {76}},\ \bibinfo {pages} {064514} (\bibinfo {year} {2007})}\BibitemShut {NoStop}%
\bibitem [{\citenamefont {Aminov}\ \emph {et~al.}(1996)\citenamefont {Aminov}, \citenamefont {Golubov},\ and\ \citenamefont {Kupriyanov}}]{aminov1996a}%
  \BibitemOpen
  \bibfield  {author} {\bibinfo {author} {\bibfnamefont {B.~A.}\ \bibnamefont {Aminov}}, \bibinfo {author} {\bibfnamefont {A.~A.}\ \bibnamefont {Golubov}}, \ and\ \bibinfo {author} {\bibfnamefont {M.~{\relax Yu}.}\ \bibnamefont {Kupriyanov}},\ }\bibfield  {title} {\enquote {\bibinfo {title} {Quasiparticle current in ballistic constrictions with finite transparencies of interfaces},}\ }\href {\doibase 10.1103/PhysRevB.53.365} {\bibfield  {journal} {\bibinfo  {journal} {Phys. Rev. B}\ }\textbf {\bibinfo {volume} {53}},\ \bibinfo {pages} {365--373} (\bibinfo {year} {1996})}\BibitemShut {NoStop}%
\bibitem [{\citenamefont {Gu{\'e}ron}\ \emph {et~al.}(1996)\citenamefont {Gu{\'e}ron}, \citenamefont {Pothier}, \citenamefont {Birge}, \citenamefont {Esteve},\ and\ \citenamefont {Devoret}}]{gueron1996}%
  \BibitemOpen
  \bibfield  {author} {\bibinfo {author} {\bibfnamefont {S.}~\bibnamefont {Gu{\'e}ron}}, \bibinfo {author} {\bibfnamefont {H.}~\bibnamefont {Pothier}}, \bibinfo {author} {\bibfnamefont {N.~O.}\ \bibnamefont {Birge}}, \bibinfo {author} {\bibfnamefont {D.}~\bibnamefont {Esteve}}, \ and\ \bibinfo {author} {\bibfnamefont {M.~H.}\ \bibnamefont {Devoret}},\ }\bibfield  {title} {\enquote {\bibinfo {title} {Superconducting {{Proximity Effect Probed}} on a {{Mesoscopic Length Scale}}},}\ }\href {\doibase 10.1103/PhysRevLett.77.3025} {\bibfield  {journal} {\bibinfo  {journal} {Phys. Rev. Lett.}\ }\textbf {\bibinfo {volume} {77}},\ \bibinfo {pages} {3025--3028} (\bibinfo {year} {1996})}\BibitemShut {NoStop}%
\bibitem [{\citenamefont {{le Sueur}}\ \emph {et~al.}(2008)\citenamefont {{le Sueur}}, \citenamefont {Joyez}, \citenamefont {Pothier}, \citenamefont {Urbina},\ and\ \citenamefont {Esteve}}]{lesueur2008}%
  \BibitemOpen
  \bibfield  {author} {\bibinfo {author} {\bibfnamefont {H.}~\bibnamefont {{le Sueur}}}, \bibinfo {author} {\bibfnamefont {P.}~\bibnamefont {Joyez}}, \bibinfo {author} {\bibfnamefont {H.}~\bibnamefont {Pothier}}, \bibinfo {author} {\bibfnamefont {C.}~\bibnamefont {Urbina}}, \ and\ \bibinfo {author} {\bibfnamefont {D.}~\bibnamefont {Esteve}},\ }\bibfield  {title} {\enquote {\bibinfo {title} {Phase {{Controlled Superconducting Proximity Effect Probed}} by {{Tunneling Spectroscopy}}},}\ }\href {\doibase 10.1103/PhysRevLett.100.197002} {\bibfield  {journal} {\bibinfo  {journal} {Phys. Rev. Lett.}\ }\textbf {\bibinfo {volume} {100}},\ \bibinfo {pages} {197002} (\bibinfo {year} {2008})}\BibitemShut {NoStop}%
\bibitem [{\citenamefont {Natterer}\ \emph {et~al.}(2016)\citenamefont {Natterer}, \citenamefont {Ha}, \citenamefont {Baek}, \citenamefont {Zhang}, \citenamefont {Cullen}, \citenamefont {Zhitenev}, \citenamefont {Kuk},\ and\ \citenamefont {Stroscio}}]{natterer2016}%
  \BibitemOpen
  \bibfield  {author} {\bibinfo {author} {\bibfnamefont {F.~D.}\ \bibnamefont {Natterer}}, \bibinfo {author} {\bibfnamefont {J.}~\bibnamefont {Ha}}, \bibinfo {author} {\bibfnamefont {H.}~\bibnamefont {Baek}}, \bibinfo {author} {\bibfnamefont {D.}~\bibnamefont {Zhang}}, \bibinfo {author} {\bibfnamefont {W.~G.}\ \bibnamefont {Cullen}}, \bibinfo {author} {\bibfnamefont {N.~B.}\ \bibnamefont {Zhitenev}}, \bibinfo {author} {\bibfnamefont {Y.}~\bibnamefont {Kuk}}, \ and\ \bibinfo {author} {\bibfnamefont {J.~A.}\ \bibnamefont {Stroscio}},\ }\bibfield  {title} {\enquote {\bibinfo {title} {Scanning tunneling spectroscopy of proximity superconductivity in epitaxial multilayer graphene},}\ }\href {\doibase 10.1103/PhysRevB.93.045406} {\bibfield  {journal} {\bibinfo  {journal} {Phys. Rev. B}\ }\textbf {\bibinfo {volume} {93}},\ \bibinfo {pages} {045406} (\bibinfo {year} {2016})}\BibitemShut {NoStop}%
\bibitem [{\citenamefont {Cort{\'e}s-del R{\'i}o}\ \emph {et~al.}(2021)\citenamefont {Cort{\'e}s-del R{\'i}o}, \citenamefont {Lado}, \citenamefont {Cherkez}, \citenamefont {Mallet}, \citenamefont {Veuillen}, \citenamefont {Cuevas}, \citenamefont {G{\'o}mez-Rodr{\'i}guez}, \citenamefont {Fern{\'a}ndez-Rossier},\ and\ \citenamefont {Brihuega}}]{cortes-delrio2021}%
  \BibitemOpen
  \bibfield  {author} {\bibinfo {author} {\bibfnamefont {E.}~\bibnamefont {Cort{\'e}s-del R{\'i}o}}, \bibinfo {author} {\bibfnamefont {J.~L.}\ \bibnamefont {Lado}}, \bibinfo {author} {\bibfnamefont {V.}~\bibnamefont {Cherkez}}, \bibinfo {author} {\bibfnamefont {P.}~\bibnamefont {Mallet}}, \bibinfo {author} {\bibfnamefont {J.-Y.}\ \bibnamefont {Veuillen}}, \bibinfo {author} {\bibfnamefont {J.~C.}\ \bibnamefont {Cuevas}}, \bibinfo {author} {\bibfnamefont {J.~M.}\ \bibnamefont {G{\'o}mez-Rodr{\'i}guez}}, \bibinfo {author} {\bibfnamefont {J.}~\bibnamefont {Fern{\'a}ndez-Rossier}}, \ and\ \bibinfo {author} {\bibfnamefont {I.}~\bibnamefont {Brihuega}},\ }\bibfield  {title} {\enquote {\bibinfo {title} {Observation of {{Yu}}--{{Shiba}}--{{Rusinov States}} in {{Superconducting Graphene}}},}\ }\href {\doibase 10.1002/adma.202008113} {\bibfield  {journal} {\bibinfo  {journal} {Adv. Mater.}\ }\textbf {\bibinfo {volume} {33}},\ \bibinfo {pages} {2008113} (\bibinfo {year} {2021})}\BibitemShut {NoStop}%
\bibitem [{\citenamefont {{Cort{\'e}s-del R{\'i}o}}\ \emph {et~al.}(2023)\citenamefont {{Cort{\'e}s-del R{\'i}o}}, \citenamefont {Trivini}, \citenamefont {Pascual}, \citenamefont {Cherkez}, \citenamefont {Mallet}, \citenamefont {Veuillen}, \citenamefont {Cuevas},\ and\ \citenamefont {Brihuega}}]{cortes-delrio2023}%
  \BibitemOpen
  \bibfield  {author} {\bibinfo {author} {\bibfnamefont {E.}~\bibnamefont {{Cort{\'e}s-del R{\'i}o}}}, \bibinfo {author} {\bibfnamefont {S.}~\bibnamefont {Trivini}}, \bibinfo {author} {\bibfnamefont {J.~I.}\ \bibnamefont {Pascual}}, \bibinfo {author} {\bibfnamefont {V.}~\bibnamefont {Cherkez}}, \bibinfo {author} {\bibfnamefont {P.}~\bibnamefont {Mallet}}, \bibinfo {author} {\bibfnamefont {J.-Y.}\ \bibnamefont {Veuillen}}, \bibinfo {author} {\bibfnamefont {J.~C.}\ \bibnamefont {Cuevas}}, \ and\ \bibinfo {author} {\bibfnamefont {I.}~\bibnamefont {Brihuega}},\ }\bibfield  {title} {\enquote {\bibinfo {title} {Shaping graphene superconductivity with nanometer precision},}\ }\href {\doibase 10.1002/smll.202308439} {\bibfield  {journal} {\bibinfo  {journal} {Small}\ }\textbf {\bibinfo {volume} {20}},\ \bibinfo {pages} {2308439} (\bibinfo {year} {2023})}\BibitemShut {NoStop}%
\bibitem [{\citenamefont {Larkin}\ and\ \citenamefont {Ovchinnikov}(1969)}]{larkin1969quasiclassical}%
  \BibitemOpen
  \bibfield  {author} {\bibinfo {author} {\bibfnamefont {A.~I.}\ \bibnamefont {Larkin}}\ and\ \bibinfo {author} {\bibfnamefont {Y.~N.}\ \bibnamefont {Ovchinnikov}},\ }\bibfield  {title} {\enquote {\bibinfo {title} {Quasiclassical method in the theory of superconductivity},}\ }\href {\doibase http://jetp.ras.ru/cgi-bin/e/index/e/28/6/p1200?a=list} {\bibfield  {journal} {\bibinfo  {journal} {Sov Phys JETP}\ }\textbf {\bibinfo {volume} {28}},\ \bibinfo {pages} {1200--1205} (\bibinfo {year} {1969})}\BibitemShut {NoStop}%
\bibitem [{\citenamefont {Eilenberger}(1968)}]{eilenberger1968transformation}%
  \BibitemOpen
  \bibfield  {author} {\bibinfo {author} {\bibfnamefont {G.}~\bibnamefont {Eilenberger}},\ }\bibfield  {title} {\enquote {\bibinfo {title} {Transformation of {{Gorkov}}'s equation for type {{II}} superconductors into transport-like equations},}\ }\href {\doibase 10.1007/BF01379803} {\bibfield  {journal} {\bibinfo  {journal} {Zeitschrift f{\"u}r Physik A Hadrons and nuclei}\ }\textbf {\bibinfo {volume} {214}},\ \bibinfo {pages} {195--213} (\bibinfo {year} {1968})}\BibitemShut {NoStop}%
\bibitem [{\citenamefont {Golubov}\ and\ \citenamefont {Kupriyanov}(1988)}]{golubov1988}%
  \BibitemOpen
  \bibfield  {author} {\bibinfo {author} {\bibfnamefont {A.~A.}\ \bibnamefont {Golubov}}\ and\ \bibinfo {author} {\bibfnamefont {M.~{\relax Yu}.}\ \bibnamefont {Kupriyanov}},\ }\bibfield  {title} {\enquote {\bibinfo {title} {Theoretical investigation of {{Josephson}} tunnel junctions with spatially inhomogeneous superconducting electrodes},}\ }\href {\doibase 10.1007/BF00683247} {\bibfield  {journal} {\bibinfo  {journal} {Journal of Low Temperature Physics}\ }\textbf {\bibinfo {volume} {70}},\ \bibinfo {pages} {83--130} (\bibinfo {year} {1988})}\BibitemShut {NoStop}%
\bibitem [{\citenamefont {Virtanen}\ and\ \citenamefont {Heikkil{\"a}}(2007)}]{virtanen2007}%
  \BibitemOpen
  \bibfield  {author} {\bibinfo {author} {\bibfnamefont {P.}~\bibnamefont {Virtanen}}\ and\ \bibinfo {author} {\bibfnamefont {T.}~\bibnamefont {Heikkil{\"a}}},\ }\bibfield  {title} {\enquote {\bibinfo {title} {Thermoelectric effects in superconducting proximity structures},}\ }\href@noop {} {\bibfield  {journal} {\bibinfo  {journal} {Appl. Phys. A}\ }\textbf {\bibinfo {volume} {89}},\ \bibinfo {pages} {625} (\bibinfo {year} {2007})},\ \bibinfo {note} {source code available at \href{https://gitlab.jyu.fi/jyucmt/usadel1}{https://gitlab.jyu.fi/jyucmt/usadel1}}\BibitemShut {NoStop}%
\bibitem [{\citenamefont {Norimatsu}\ and\ \citenamefont {Kusunoki}(2014)}]{norimatsu2014}%
  \BibitemOpen
  \bibfield  {author} {\bibinfo {author} {\bibfnamefont {W.}~\bibnamefont {Norimatsu}}\ and\ \bibinfo {author} {\bibfnamefont {M.}~\bibnamefont {Kusunoki}},\ }\bibfield  {title} {\enquote {\bibinfo {title} {Epitaxial graphene on {{SiC}}\{0001\}: Advances and perspectives},}\ }\href {\doibase 10.1039/c3cp54523g} {\bibfield  {journal} {\bibinfo  {journal} {Phys. Chem. Chem. Phys.}\ }\textbf {\bibinfo {volume} {16}},\ \bibinfo {pages} {3501} (\bibinfo {year} {2014})}\BibitemShut {NoStop}%
\bibitem [{SI()}]{SI}%
  \BibitemOpen
  \href@noop {} {}\bibinfo {note} {See Supplemental Material at [URL] for details on the model and additional data.}\BibitemShut {Stop}%
\bibitem [{\citenamefont {{Luxmi}}\ \emph {et~al.}(2010)\citenamefont {{Luxmi}}, \citenamefont {Srivastava}, \citenamefont {He}, \citenamefont {Feenstra},\ and\ \citenamefont {Fisher}}]{luxmi2010}%
  \BibitemOpen
  \bibfield  {author} {\bibinfo {author} {\bibnamefont {{Luxmi}}}, \bibinfo {author} {\bibfnamefont {N.}~\bibnamefont {Srivastava}}, \bibinfo {author} {\bibfnamefont {G.}~\bibnamefont {He}}, \bibinfo {author} {\bibfnamefont {R.~M.}\ \bibnamefont {Feenstra}}, \ and\ \bibinfo {author} {\bibfnamefont {P.~J.}\ \bibnamefont {Fisher}},\ }\bibfield  {title} {\enquote {\bibinfo {title} {Comparison of graphene formation on {{C-face}} and {{Si-face SiC}} \{0001\} surfaces},}\ }\href {\doibase 10.1103/PhysRevB.82.235406} {\bibfield  {journal} {\bibinfo  {journal} {Phys. Rev. B}\ }\textbf {\bibinfo {volume} {82}},\ \bibinfo {pages} {235406} (\bibinfo {year} {2010})}\BibitemShut {NoStop}%
\bibitem [{\citenamefont {Rodrigo}\ \emph {et~al.}(2004)\citenamefont {Rodrigo}, \citenamefont {Suderow},\ and\ \citenamefont {Vieira}}]{rodrigo2004}%
  \BibitemOpen
  \bibfield  {author} {\bibinfo {author} {\bibfnamefont {J.~G.}\ \bibnamefont {Rodrigo}}, \bibinfo {author} {\bibfnamefont {H.}~\bibnamefont {Suderow}}, \ and\ \bibinfo {author} {\bibfnamefont {S.}~\bibnamefont {Vieira}},\ }\bibfield  {title} {\enquote {\bibinfo {title} {On the use of {{STM}} superconducting tips at very low temperatures},}\ }\href {\doibase 10.1140/epjb/e2004-00273-y} {\bibfield  {journal} {\bibinfo  {journal} {The European Physical Journal B - Condensed Matter and Complex Systems}\ }\textbf {\bibinfo {volume} {40}},\ \bibinfo {pages} {483--488} (\bibinfo {year} {2004})}\BibitemShut {NoStop}%
\bibitem [{\citenamefont {Franke}\ \emph {et~al.}(2011)\citenamefont {Franke}, \citenamefont {Schulze},\ and\ \citenamefont {Pascual}}]{franke2011a}%
  \BibitemOpen
  \bibfield  {author} {\bibinfo {author} {\bibfnamefont {K.~J.}\ \bibnamefont {Franke}}, \bibinfo {author} {\bibfnamefont {G.}~\bibnamefont {Schulze}}, \ and\ \bibinfo {author} {\bibfnamefont {J.~I.}\ \bibnamefont {Pascual}},\ }\bibfield  {title} {\enquote {\bibinfo {title} {Competition of {{Superconducting Phenomena}} and {{Kondo Screening}} at the {{Nanoscale}}},}\ }\href {\doibase 10.1126/science.1202204} {\bibfield  {journal} {\bibinfo  {journal} {Science}\ }\textbf {\bibinfo {volume} {332}},\ \bibinfo {pages} {940--944} (\bibinfo {year} {2011})}\BibitemShut {NoStop}%
\bibitem [{\citenamefont {Pillet}\ \emph {et~al.}(2010)\citenamefont {Pillet}, \citenamefont {Quay}, \citenamefont {Morfin}, \citenamefont {Bena}, \citenamefont {Yeyati},\ and\ \citenamefont {Joyez}}]{pillet2010a}%
  \BibitemOpen
  \bibfield  {author} {\bibinfo {author} {\bibfnamefont {J.-D.}\ \bibnamefont {Pillet}}, \bibinfo {author} {\bibfnamefont {C.~H.~L.}\ \bibnamefont {Quay}}, \bibinfo {author} {\bibfnamefont {P.}~\bibnamefont {Morfin}}, \bibinfo {author} {\bibfnamefont {C.}~\bibnamefont {Bena}}, \bibinfo {author} {\bibfnamefont {A.~L.}\ \bibnamefont {Yeyati}}, \ and\ \bibinfo {author} {\bibfnamefont {P.}~\bibnamefont {Joyez}},\ }\bibfield  {title} {\enquote {\bibinfo {title} {Andreev bound states in supercurrent-carrying carbon nanotubes revealed},}\ }\href {\doibase 10.1038/nphys1811} {\bibfield  {journal} {\bibinfo  {journal} {Nature Phys}\ }\textbf {\bibinfo {volume} {6}},\ \bibinfo {pages} {965--969} (\bibinfo {year} {2010})}\BibitemShut {NoStop}%
\bibitem [{\citenamefont {Ruby}\ \emph {et~al.}(2015)\citenamefont {Ruby}, \citenamefont {Heinrich}, \citenamefont {Pascual},\ and\ \citenamefont {Franke}}]{ruby2015}%
  \BibitemOpen
  \bibfield  {author} {\bibinfo {author} {\bibfnamefont {M.}~\bibnamefont {Ruby}}, \bibinfo {author} {\bibfnamefont {B.~W.}\ \bibnamefont {Heinrich}}, \bibinfo {author} {\bibfnamefont {J.~I.}\ \bibnamefont {Pascual}}, \ and\ \bibinfo {author} {\bibfnamefont {K.~J.}\ \bibnamefont {Franke}},\ }\bibfield  {title} {\enquote {\bibinfo {title} {Experimental demonstration of a two-band superconducting state for lead using scanning tunneling spectroscopy},}\ }\href {\doibase 10.1103/PhysRevLett.114.157001} {\bibfield  {journal} {\bibinfo  {journal} {Physical Review Letters}\ }\textbf {\bibinfo {volume} {114}},\ \bibinfo {pages} {157001} (\bibinfo {year} {2015})}\BibitemShut {NoStop}%
\bibitem [{\citenamefont {Hass}\ \emph {et~al.}(2008)\citenamefont {Hass}, \citenamefont {Varchon}, \citenamefont {{Mill{\'a}n-Otoya}}, \citenamefont {Sprinkle}, \citenamefont {Sharma}, \citenamefont {{de Heer}}, \citenamefont {Berger}, \citenamefont {First}, \citenamefont {Magaud},\ and\ \citenamefont {Conrad}}]{hass2008}%
  \BibitemOpen
  \bibfield  {author} {\bibinfo {author} {\bibfnamefont {J.}~\bibnamefont {Hass}}, \bibinfo {author} {\bibfnamefont {F.}~\bibnamefont {Varchon}}, \bibinfo {author} {\bibfnamefont {J.~E.}\ \bibnamefont {{Mill{\'a}n-Otoya}}}, \bibinfo {author} {\bibfnamefont {M.}~\bibnamefont {Sprinkle}}, \bibinfo {author} {\bibfnamefont {N.}~\bibnamefont {Sharma}}, \bibinfo {author} {\bibfnamefont {W.~A.}\ \bibnamefont {{de Heer}}}, \bibinfo {author} {\bibfnamefont {C.}~\bibnamefont {Berger}}, \bibinfo {author} {\bibfnamefont {P.~N.}\ \bibnamefont {First}}, \bibinfo {author} {\bibfnamefont {L.}~\bibnamefont {Magaud}}, \ and\ \bibinfo {author} {\bibfnamefont {E.~H.}\ \bibnamefont {Conrad}},\ }\bibfield  {title} {\enquote {\bibinfo {title} {Why {{Multilayer Graphene}} on 4 {{H}} - {{SiC}} ( 000 1 {\textasciimacron} ) {{Behaves Like}} a {{Single Sheet}} of {{Graphene}}},}\ }\href {\doibase 10.1103/PhysRevLett.100.125504} {\bibfield  {journal} {\bibinfo  {journal} {Phys. Rev. Lett.}\ }\textbf {\bibinfo {volume} {100}},\ \bibinfo
  {pages} {125504} (\bibinfo {year} {2008})}\BibitemShut {NoStop}%
\bibitem [{\citenamefont {Sprinkle}\ \emph {et~al.}(2009)\citenamefont {Sprinkle}, \citenamefont {Siegel}, \citenamefont {Hu}, \citenamefont {Hicks}, \citenamefont {Tejeda}, \citenamefont {{Taleb-Ibrahimi}}, \citenamefont {Le~F{\`e}vre}, \citenamefont {Bertran}, \citenamefont {Vizzini}, \citenamefont {Enriquez}, \citenamefont {Chiang}, \citenamefont {Soukiassian}, \citenamefont {Berger}, \citenamefont {{de Heer}}, \citenamefont {Lanzara},\ and\ \citenamefont {Conrad}}]{sprinkle2009}%
  \BibitemOpen
  \bibfield  {author} {\bibinfo {author} {\bibfnamefont {M.}~\bibnamefont {Sprinkle}}, \bibinfo {author} {\bibfnamefont {D.}~\bibnamefont {Siegel}}, \bibinfo {author} {\bibfnamefont {Y.}~\bibnamefont {Hu}}, \bibinfo {author} {\bibfnamefont {J.}~\bibnamefont {Hicks}}, \bibinfo {author} {\bibfnamefont {A.}~\bibnamefont {Tejeda}}, \bibinfo {author} {\bibfnamefont {A.}~\bibnamefont {{Taleb-Ibrahimi}}}, \bibinfo {author} {\bibfnamefont {P.}~\bibnamefont {Le~F{\`e}vre}}, \bibinfo {author} {\bibfnamefont {F.}~\bibnamefont {Bertran}}, \bibinfo {author} {\bibfnamefont {S.}~\bibnamefont {Vizzini}}, \bibinfo {author} {\bibfnamefont {H.}~\bibnamefont {Enriquez}}, \bibinfo {author} {\bibfnamefont {S.}~\bibnamefont {Chiang}}, \bibinfo {author} {\bibfnamefont {P.}~\bibnamefont {Soukiassian}}, \bibinfo {author} {\bibfnamefont {C.}~\bibnamefont {Berger}}, \bibinfo {author} {\bibfnamefont {W.~A.}\ \bibnamefont {{de Heer}}}, \bibinfo {author} {\bibfnamefont {A.}~\bibnamefont {Lanzara}}, \ and\ \bibinfo {author} {\bibfnamefont
  {E.~H.}\ \bibnamefont {Conrad}},\ }\bibfield  {title} {\enquote {\bibinfo {title} {First direct observation of a nearly ideal graphene band structure},}\ }\href {\doibase 10.1103/PhysRevLett.103.226803} {\bibfield  {journal} {\bibinfo  {journal} {Phys. Rev. Lett.}\ }\textbf {\bibinfo {volume} {103}},\ \bibinfo {pages} {226803} (\bibinfo {year} {2009})}\BibitemShut {NoStop}%
\bibitem [{\citenamefont {Varchon}\ \emph {et~al.}(2008)\citenamefont {Varchon}, \citenamefont {Mallet}, \citenamefont {Magaud},\ and\ \citenamefont {Veuillen}}]{varchon2008}%
  \BibitemOpen
  \bibfield  {author} {\bibinfo {author} {\bibfnamefont {F.}~\bibnamefont {Varchon}}, \bibinfo {author} {\bibfnamefont {P.}~\bibnamefont {Mallet}}, \bibinfo {author} {\bibfnamefont {L.}~\bibnamefont {Magaud}}, \ and\ \bibinfo {author} {\bibfnamefont {J.-Y.}\ \bibnamefont {Veuillen}},\ }\bibfield  {title} {\enquote {\bibinfo {title} {Rotational disorder in few-layer graphene films on 6 {{H}} - {{Si C}} ( 000 - 1 ) : {{A}} scanning tunneling microscopy study},}\ }\href {\doibase 10.1103/PhysRevB.77.165415} {\bibfield  {journal} {\bibinfo  {journal} {Phys. Rev. B}\ }\textbf {\bibinfo {volume} {77}},\ \bibinfo {pages} {165415} (\bibinfo {year} {2008})}\BibitemShut {NoStop}%
\bibitem [{\citenamefont {Eigler}\ and\ \citenamefont {Schweizer}(1990)}]{eigler1990}%
  \BibitemOpen
  \bibfield  {author} {\bibinfo {author} {\bibfnamefont {D.~M.}\ \bibnamefont {Eigler}}\ and\ \bibinfo {author} {\bibfnamefont {E.~K.}\ \bibnamefont {Schweizer}},\ }\bibfield  {title} {\enquote {\bibinfo {title} {Positioning single atoms with a scanning tunnelling microscope},}\ }\href {\doibase 10.1038/344524a0} {\bibfield  {journal} {\bibinfo  {journal} {Nature}\ }\textbf {\bibinfo {volume} {344}},\ \bibinfo {pages} {524--526} (\bibinfo {year} {1990})}\BibitemShut {NoStop}%
\bibitem [{\citenamefont {Schneider}\ \emph {et~al.}(2023)\citenamefont {Schneider}, \citenamefont {Ton}, \citenamefont {Ioannidis}, \citenamefont {{Neuhaus-Steinmetz}}, \citenamefont {Posske}, \citenamefont {Wiesendanger},\ and\ \citenamefont {Wiebe}}]{schneider2023}%
  \BibitemOpen
  \bibfield  {author} {\bibinfo {author} {\bibfnamefont {L.}~\bibnamefont {Schneider}}, \bibinfo {author} {\bibfnamefont {K.~T.}\ \bibnamefont {Ton}}, \bibinfo {author} {\bibfnamefont {I.}~\bibnamefont {Ioannidis}}, \bibinfo {author} {\bibfnamefont {J.}~\bibnamefont {{Neuhaus-Steinmetz}}}, \bibinfo {author} {\bibfnamefont {T.}~\bibnamefont {Posske}}, \bibinfo {author} {\bibfnamefont {R.}~\bibnamefont {Wiesendanger}}, \ and\ \bibinfo {author} {\bibfnamefont {J.}~\bibnamefont {Wiebe}},\ }\bibfield  {title} {\enquote {\bibinfo {title} {Proximity superconductivity in atom-by-atom crafted quantum dots},}\ }\href {\doibase 10.1038/s41586-023-06312-0} {\bibfield  {journal} {\bibinfo  {journal} {Nature}\ }\textbf {\bibinfo {volume} {621}},\ \bibinfo {pages} {60--65} (\bibinfo {year} {2023})}\BibitemShut {NoStop}%
\bibitem [{\citenamefont {R{\"u}tten}\ \emph {et~al.}(2024)\citenamefont {R{\"u}tten}, \citenamefont {Schmid}, \citenamefont {Liebhaber}, \citenamefont {Franceschi}, \citenamefont {Yazdani}, \citenamefont {Reecht}, \citenamefont {Rossnagel}, \citenamefont {{von Oppen}},\ and\ \citenamefont {Franke}}]{rutten2024a}%
  \BibitemOpen
  \bibfield  {author} {\bibinfo {author} {\bibfnamefont {L.~M.}\ \bibnamefont {R{\"u}tten}}, \bibinfo {author} {\bibfnamefont {H.}~\bibnamefont {Schmid}}, \bibinfo {author} {\bibfnamefont {E.}~\bibnamefont {Liebhaber}}, \bibinfo {author} {\bibfnamefont {G.}~\bibnamefont {Franceschi}}, \bibinfo {author} {\bibfnamefont {A.}~\bibnamefont {Yazdani}}, \bibinfo {author} {\bibfnamefont {G.}~\bibnamefont {Reecht}}, \bibinfo {author} {\bibfnamefont {K.}~\bibnamefont {Rossnagel}}, \bibinfo {author} {\bibfnamefont {F.}~\bibnamefont {{von Oppen}}}, \ and\ \bibinfo {author} {\bibfnamefont {K.~J.}\ \bibnamefont {Franke}},\ }\bibfield  {title} {\enquote {\bibinfo {title} {Wave {{Function Engineering}} on {{Superconducting Substrates}}: {{Chiral Yu-Shiba-Rusinov Molecules}}},}\ }\href {\doibase 10.1021/acsnano.4c10998} {\bibfield  {journal} {\bibinfo  {journal} {ACS Nano}\ }\textbf {\bibinfo {volume} {18}},\ \bibinfo {pages} {30798--30804} (\bibinfo {year} {2024})}\BibitemShut {NoStop}%
\bibitem [{\citenamefont {Trivini}\ \emph {et~al.}(2024)\citenamefont {Trivini}, \citenamefont {Ortuzar}, \citenamefont {Zaldivar}, \citenamefont {Herrera}, \citenamefont {Guillam{\'o}n}, \citenamefont {Suderow}, \citenamefont {Bergeret},\ and\ \citenamefont {Pascual}}]{trivini2024b}%
  \BibitemOpen
  \bibfield  {author} {\bibinfo {author} {\bibfnamefont {S.}~\bibnamefont {Trivini}}, \bibinfo {author} {\bibfnamefont {J.}~\bibnamefont {Ortuzar}}, \bibinfo {author} {\bibfnamefont {J.}~\bibnamefont {Zaldivar}}, \bibinfo {author} {\bibfnamefont {E.}~\bibnamefont {Herrera}}, \bibinfo {author} {\bibfnamefont {I.}~\bibnamefont {Guillam{\'o}n}}, \bibinfo {author} {\bibfnamefont {H.}~\bibnamefont {Suderow}}, \bibinfo {author} {\bibfnamefont {F.~S.}\ \bibnamefont {Bergeret}}, \ and\ \bibinfo {author} {\bibfnamefont {J.~I.}\ \bibnamefont {Pascual}},\ }\bibfield  {title} {\enquote {\bibinfo {title} {Diluted {{Yu-Shiba-Rusinov}} arrays on the anisotropic superconductor {$\beta$}-{{Bi}}{$_{2}$}{{Pd}}},}\ }\href {\doibase 10.1103/PhysRevB.110.235405} {\bibfield  {journal} {\bibinfo  {journal} {Phys. Rev. B}\ }\textbf {\bibinfo {volume} {110}},\ \bibinfo {pages} {235405} (\bibinfo {year} {2024})}\BibitemShut {NoStop}%
\bibitem [{\citenamefont {Feigel'man}\ \emph {et~al.}(2008)\citenamefont {Feigel'man}, \citenamefont {Skvortsov},\ and\ \citenamefont {Tikhonov}}]{feigelman2008}%
  \BibitemOpen
  \bibfield  {author} {\bibinfo {author} {\bibfnamefont {M.~V.}\ \bibnamefont {Feigel'man}}, \bibinfo {author} {\bibfnamefont {M.~A.}\ \bibnamefont {Skvortsov}}, \ and\ \bibinfo {author} {\bibfnamefont {K.~S.}\ \bibnamefont {Tikhonov}},\ }\bibfield  {title} {\enquote {\bibinfo {title} {Proximity-induced superconductivity in graphene},}\ }\href {\doibase 10.1134/S0021364008230100} {\bibfield  {journal} {\bibinfo  {journal} {JETP Lett.}\ }\textbf {\bibinfo {volume} {88}},\ \bibinfo {pages} {747--751} (\bibinfo {year} {2008})}\BibitemShut {NoStop}%
\bibitem [{\citenamefont {Kessler}\ \emph {et~al.}(2010)\citenamefont {Kessler}, \citenamefont {Girit}, \citenamefont {Zettl},\ and\ \citenamefont {Bouchiat}}]{kessler2010a}%
  \BibitemOpen
  \bibfield  {author} {\bibinfo {author} {\bibfnamefont {B.~M.}\ \bibnamefont {Kessler}}, \bibinfo {author} {\bibfnamefont {{\c C}.~{\"O}.}\ \bibnamefont {Girit}}, \bibinfo {author} {\bibfnamefont {A.}~\bibnamefont {Zettl}}, \ and\ \bibinfo {author} {\bibfnamefont {V.}~\bibnamefont {Bouchiat}},\ }\bibfield  {title} {\enquote {\bibinfo {title} {Tunable {{Superconducting Phase Transition}} in {{Metal-Decorated Graphene Sheets}}},}\ }\href {\doibase 10.1103/PhysRevLett.104.047001} {\bibfield  {journal} {\bibinfo  {journal} {Phys. Rev. Lett.}\ }\textbf {\bibinfo {volume} {104}},\ \bibinfo {pages} {047001} (\bibinfo {year} {2010})}\BibitemShut {NoStop}%
\bibitem [{\citenamefont {Eley}\ \emph {et~al.}(2012)\citenamefont {Eley}, \citenamefont {Gopalakrishnan}, \citenamefont {Goldbart},\ and\ \citenamefont {Mason}}]{eley2012}%
  \BibitemOpen
  \bibfield  {author} {\bibinfo {author} {\bibfnamefont {S.}~\bibnamefont {Eley}}, \bibinfo {author} {\bibfnamefont {S.}~\bibnamefont {Gopalakrishnan}}, \bibinfo {author} {\bibfnamefont {P.~M.}\ \bibnamefont {Goldbart}}, \ and\ \bibinfo {author} {\bibfnamefont {N.}~\bibnamefont {Mason}},\ }\bibfield  {title} {\enquote {\bibinfo {title} {Approaching zero-temperature metallic states in mesoscopic superconductor--normal--superconductor arrays},}\ }\href {\doibase 10.1038/nphys2154} {\bibfield  {journal} {\bibinfo  {journal} {Nat. Phys.}\ }\textbf {\bibinfo {volume} {8}},\ \bibinfo {pages} {59--62} (\bibinfo {year} {2012})}\BibitemShut {NoStop}%
\bibitem [{\citenamefont {Chen}\ \emph {et~al.}(2024)\citenamefont {Chen}, \citenamefont {Liu}, \citenamefont {Guo}, \citenamefont {Wang}, \citenamefont {Tian}, \citenamefont {Zhang}, \citenamefont {Xue}, \citenamefont {Mu}, \citenamefont {Zhang},\ and\ \citenamefont {Di}}]{chen2024}%
  \BibitemOpen
  \bibfield  {author} {\bibinfo {author} {\bibfnamefont {F.}~\bibnamefont {Chen}}, \bibinfo {author} {\bibfnamefont {Y.}~\bibnamefont {Liu}}, \bibinfo {author} {\bibfnamefont {W.}~\bibnamefont {Guo}}, \bibinfo {author} {\bibfnamefont {T.}~\bibnamefont {Wang}}, \bibinfo {author} {\bibfnamefont {Z.}~\bibnamefont {Tian}}, \bibinfo {author} {\bibfnamefont {M.}~\bibnamefont {Zhang}}, \bibinfo {author} {\bibfnamefont {Z.}~\bibnamefont {Xue}}, \bibinfo {author} {\bibfnamefont {G.}~\bibnamefont {Mu}}, \bibinfo {author} {\bibfnamefont {X.}~\bibnamefont {Zhang}}, \ and\ \bibinfo {author} {\bibfnamefont {Z.}~\bibnamefont {Di}},\ }\bibfield  {title} {\enquote {\bibinfo {title} {Quantum {{Griffiths Singularity}} in {{Ordered Artificial Superconducting-Islands-Array}} on {{Graphene}}},}\ }\href {\doibase 10.1021/acs.nanolett.3c03870} {\bibfield  {journal} {\bibinfo  {journal} {Nano Lett.}\ }\textbf {\bibinfo {volume} {24}},\ \bibinfo {pages} {2444--2450} (\bibinfo {year} {2024})}\BibitemShut {NoStop}%
\bibitem [{\citenamefont {Mammadov}\ \emph {et~al.}(2017)\citenamefont {Mammadov}, \citenamefont {Ristein}, \citenamefont {Krone}, \citenamefont {Raidel}, \citenamefont {Wanke}, \citenamefont {Wiesmann}, \citenamefont {Speck},\ and\ \citenamefont {Seyller}}]{Mammadov2017}%
  \BibitemOpen
  \bibfield  {author} {\bibinfo {author} {\bibfnamefont {S.}~\bibnamefont {Mammadov}}, \bibinfo {author} {\bibfnamefont {J.}~\bibnamefont {Ristein}}, \bibinfo {author} {\bibfnamefont {J.}~\bibnamefont {Krone}}, \bibinfo {author} {\bibfnamefont {C.}~\bibnamefont {Raidel}}, \bibinfo {author} {\bibfnamefont {M.}~\bibnamefont {Wanke}}, \bibinfo {author} {\bibfnamefont {V.}~\bibnamefont {Wiesmann}}, \bibinfo {author} {\bibfnamefont {F.}~\bibnamefont {Speck}}, \ and\ \bibinfo {author} {\bibfnamefont {T.}~\bibnamefont {Seyller}},\ }\bibfield  {title} {\enquote {\bibinfo {title} {Work function of graphene multilayers on {{SiC}}(0001)},}\ }\href {\doibase 10.1088/2053-1583/4/1/015043} {\bibfield  {journal} {\bibinfo  {journal} {2D Mater.}\ }\textbf {\bibinfo {volume} {4}},\ \bibinfo {pages} {015043} (\bibinfo {year} {2017})}\BibitemShut {NoStop}%
\bibitem [{\citenamefont {Giovannetti}\ \emph {et~al.}(2008)\citenamefont {Giovannetti}, \citenamefont {Khomyakov}, \citenamefont {Brocks}, \citenamefont {Karpan}, \citenamefont {Van Den~Brink},\ and\ \citenamefont {Kelly}}]{Giovannetti2008}%
  \BibitemOpen
  \bibfield  {author} {\bibinfo {author} {\bibfnamefont {G.}~\bibnamefont {Giovannetti}}, \bibinfo {author} {\bibfnamefont {P.~A.}\ \bibnamefont {Khomyakov}}, \bibinfo {author} {\bibfnamefont {G.}~\bibnamefont {Brocks}}, \bibinfo {author} {\bibfnamefont {V.~M.}\ \bibnamefont {Karpan}}, \bibinfo {author} {\bibfnamefont {J.}~\bibnamefont {Van Den~Brink}}, \ and\ \bibinfo {author} {\bibfnamefont {P.~J.}\ \bibnamefont {Kelly}},\ }\bibfield  {title} {\enquote {\bibinfo {title} {Doping {{Graphene}} with {{Metal Contacts}}},}\ }\href {\doibase 10.1103/PhysRevLett.101.026803} {\bibfield  {journal} {\bibinfo  {journal} {Phys. Rev. Lett.}\ }\textbf {\bibinfo {volume} {101}},\ \bibinfo {pages} {026803} (\bibinfo {year} {2008})}\BibitemShut {NoStop}%
\bibitem [{\citenamefont {Han}\ \emph {et~al.}(2014)\citenamefont {Han}, \citenamefont {Allain}, \citenamefont {{Arjmandi-Tash}}, \citenamefont {Tikhonov}, \citenamefont {Feigel'man}, \citenamefont {Sac{\'e}p{\'e}},\ and\ \citenamefont {Bouchiat}}]{han2014a}%
  \BibitemOpen
  \bibfield  {author} {\bibinfo {author} {\bibfnamefont {Z.}~\bibnamefont {Han}}, \bibinfo {author} {\bibfnamefont {A.}~\bibnamefont {Allain}}, \bibinfo {author} {\bibfnamefont {H.}~\bibnamefont {{Arjmandi-Tash}}}, \bibinfo {author} {\bibfnamefont {K.}~\bibnamefont {Tikhonov}}, \bibinfo {author} {\bibfnamefont {M.}~\bibnamefont {Feigel'man}}, \bibinfo {author} {\bibfnamefont {B.}~\bibnamefont {Sac{\'e}p{\'e}}}, \ and\ \bibinfo {author} {\bibfnamefont {V.}~\bibnamefont {Bouchiat}},\ }\bibfield  {title} {\enquote {\bibinfo {title} {Collapse of superconductivity in a hybrid tin--graphene {{Josephson}} junction array},}\ }\href@noop {} {\bibfield  {journal} {\bibinfo  {journal} {Nat. Phys.}\ }\textbf {\bibinfo {volume} {10}},\ \bibinfo {pages} {380--386} (\bibinfo {year} {2014})}\BibitemShut {NoStop}%
\end{thebibliography}

%

\end{document}